\documentclass[preprint,12pt]{elsarticle}

\usepackage{graphicx}

\usepackage{amssymb}

\usepackage[left,modulo]{lineno}

\journal{Nuclear Instruments and Methods A}

\begin{document}

\begin{frontmatter}



\title{Results of the BiPo-1 prototype for radiopurity measurements for the SuperNEMO double beta decay source foils}


\author[lal]{J.~Argyriades}
\author[iphc]{R.~Arnold}
\author[lal]{C.~Augier}
\author[inl]{J.~Baker}
\author[itep]{A.S.~Barabash}
\author[ucl]{A.~Basharina-Freshville}
\author[lal]{M.~Bongrand}
\author[lal]{C.~Bourgeois}
\author[lal]{D.~Breton}
\author[lal]{M.~Bri\`ere}
\author[lal]{G.~Broudin-Bay}
\author[jinr]{V.B.~Brudanin}
\author[inl]{A.J.~Caffrey}
\author[zaragoza]{S.~Cebri\'an}
\author[lpc]{A.~Chapon}
\author[cenbg1,cenbg2]{E.~Chauveau}
\author[zaragoza]{Th.~Dafni}
\author[ific]{J.~D\'{\i}az}
\author[lpc]{D.~Durand}
\author[jinr]{V.G.~Egorov}
\author[ucl]{J.J.~Evans}
\author[ucl]{R.~Flack}
\author[tokushima]{K-I.~Fushima}
\author[zaragoza]{I.G.~Irastorza}
\author[lal]{X.~Garrido}
\author[zaragoza]{H.~G\'{o}mez}
\author[lpc]{B.~Guillon}
\author[ucl]{A.~Holin}
\author[lpc]{J.~Hommet}
\author[bratislava]{K~Holy}
\author[inl]{J.J.~Horkey}
\author[cenbg1,cenbg2]{P.~Hubert}
\author[cenbg1,cenbg2]{C.~Hugon}
\author[zaragoza]{F.J.~Iguaz}
\author[kek]{N.~Ishihara}
\author[manchester]{C.~M.~Jackson}
\author[lal]{S.~Jenzer}
\author[lal]{S.~Jullian}
\author[ucl]{M.~Kauer}
\author[jinr]{O.I.~Kochetov}
\author[itep]{S.I.~Konovalov}
\author[iphc,jinr]{V.~Kovalenko}
\author[maroc]{T.~Lamhamdi}
\author[texas]{K.~Lang}
\author[lpc]{Y.~Lemi\`ere}
\author[cenbg1,cenbg2]{G.~Lutter}
\author[zaragoza]{G.~Luz\'{o}n}
\author[ieap]{F.~Mamedov}
\author[cenbg1,cenbg2]{Ch.~Marquet}
\author[lpc]{F.~Mauger}
\author[ific]{F.~Monrabal}
\author[cenbg1,cenbg2]{A.~Nachab}
\author[jinr]{I.B.~Nemchenok}
\author[cenbg1,cenbg2]{C.H.~Nguyen}
\author[osaka]{M.~Nomachi}
\author[barcelona]{F.~Nova}
\author[saga]{H.~Ohsumi}
\author[texas]{R.B.~Pahlka}
\author[cenbg1,cenbg2]{F.~Perrot}
\author[cenbg1,cenbg2]{F.~Piquemal}
\author[bratislava]{P.P.~Povinec}
\author[ucl]{B.~Richards}
\author[cenbg1,cenbg2]{J.S.~Ricol}
\author[inl]{C.L.~Riddle}
\author[zaragoza]{A.~Rodr\'{\i}guez}
\author[ucl]{R.~Saakyan}
\author[lal]{X.~Sarazin\corref{cor1}}
\author[imperial]{J.K.~Sedgbeer}
\author[ific]{L.~Serra}
\author[imperial]{Yu.A.~Shitov}
\author[lal]{L.~Simard}
\author[bratislava]{F.~\v{S}imkovic}
\author[manchester]{S.~S\"oldner-Rembold}
\author[ieap]{I.~\v{S}tekl}
\author[mhc]{C.S.~Sutton}
\author[fukumi]{Y.~Tamagawa}
\author[lal]{G.~Szklarz}
\author[ucl]{J.~Thomas}
\author[jinr]{V.~Timkin}
\author[jinr]{V.~Tretyak}
\author[kiev]{Vl.I.~Tretyak}
\author[itep]{V.I.~Umatov}
\author[ieap]{L.~V\'{a}la}
\author[itep]{I.A.~Vanyushin}
\author[jinr]{R.~Vasiliev}
\author[ucl]{V.A.~Vasiliev}
\author[charles]{V.~Vorobel}
\author[ucl]{D.~Waters}
\author[ific]{N.~Yahali}
\author[charles]{A.~\v{Z}ukauskas}

\cortext[cor1]{Corresponding authors: sarazin@lal.in2p3.fr}

\address[lal]{LAL, Universit\'e Paris-Sud, CNRS/IN2P3, F-91405 Orsay, France}
\address[iphc]{IPHC, Universit\'e de Strasbourg, CNRS/IN2P3, F-67037 Strasbourg, France}
\address[inl]{INL, Idaho Falls, ID 83415, USA}
\address[itep]{Institute of Theoretical and Experimental Physics, 117259 Moscow, Russia}
\address[ucl]{University College London, WC1E 6BT London, United Kingdom}
\address[jinr]{Joint Institute for Nuclear Research, 141980 Dubna, Russia}
\address[zaragoza]{Instituto de Fisica Nuclear y Altas Energias, Universidad de Zaragoza, Zaragoza, Spain}
\address[lpc]{LPC Caen, ENSICAEN, Universit\'e de Caen, CNRS/IN2P3, F-14032 Caen, France}
\address[cenbg1]{CNRS/IN2P3, Centre d'Etudes Nucl\'eaires de Bordeaux Gradignan, UMR 5797, F-33175 Gradignan, France}
\address[cenbg2]{Universit\'e de Bordeaux, Centre d'Etudes Nucl\'eaires de Bordeaux Gradignan, UMR 5797, F-33175 Gradignan, France}
\address[manchester]{University of Manchester, M13 9PL Manchester, United Kingdom}
\address[ific]{Instituto de Fisica Corpuscular, CSIC, Universidad de Valencia, Valencia, Spain}
\address[tokushima]{Tokushima University, Japan}
\address[bratislava]{FMFI, Comenius University, SK-842 48 Bratislava, Slovakia}
\address[kek]{KEK, Japan}
\address[maroc]{USMBA, Fes, Morocco}
\address[texas]{Department of Physics, 1 University Station C1600, The University of Texas at Austin, Austin, TX 78712, USA}
\address[ieap]{IEAP, Czech Technical University in Prague, CZ-12800 Prague, Czech Republic}
\address[osaka]{Osaka University, Osaka, Japan}
\address[barcelona]{Universitat Autonoma de Barcelona, Spain}
\address[saga]{Saga University, Saga 840-8502, Japan}
\address[imperial]{Imperial College London, United Kingdom}
\address[mhc]{MHC, South Hadley, Massachusetts, MA 01075, USA}
\address[fukumi]{Fukui University, Japan}
\address[kiev]{INR, MSP 03680 Kyiv, Ukraine}
\address[charles]{Charles University in Prague, Faculty of Mathematics and Physics, CZ-12116 Prague, Czech Republic}

\begin{abstract}
The development of BiPo detectors is dedicated to the measurement of extremely high radiopurity in $^{208}$Tl and $^{214}$Bi  for the SuperNEMO double beta decay source foils. A modular prototype, called BiPo-1, with 0.8~$m^2$ of sensitive surface area, has been running in the Modane Underground Laboratory since February, 2008. The goal of BiPo-1 is to measure the different components of the background and in particular the surface radiopurity of the plastic scintillators that make up the detector. The first phase of data collection has been dedicated to the measurement of the radiopurity in $^{208}$Tl. After more than one year of background measurement, a surface activity of the scintillators of $\mathcal{A}$($^{208}$Tl)~$=$~1.5~$\mu$Bq/m$^2$ is reported here. 
Given this level of background, a larger BiPo detector having 12~m$^2$ of active surface area, is able to qualify the radiopurity of the SuperNEMO selenium double beta decay foils with the required sensitivity of $\mathcal{A}$($^{208}$Tl)~$<$~2~$\mu$Bq/kg (90\% C.L.) with a six month measurement.
\end{abstract}

\begin{keyword}

Double beta decay \sep NEMO-3 \sep SuperNEMO \sep BiPo \sep Neutrino \sep Majorana \sep Radiopurity

\end{keyword}

\end{frontmatter}

\linenumbers


\section*{Introduction}
\label{sec:introduction}

The search for neutrinoless double beta decay ($\beta\beta 0\nu$),  is a major challenge in particle physics. The observation of this decay, a process beyond the Standard Model which violates lepton number by two units, would be an experimental proof that the neutrino is a Majorana particle, i.e. identical to its antiparticle. 
The NEMO-3 detector~\cite{nemo3a,nemo3b,nemo3c,nemo3d,nemo3e,nemo3f,nemo3g} has been running since 2003 in the Modane Underground Laboratory and is devoted to the search of $\beta\beta 0\nu$ decay. NEMO-3 uses a combination of a tracking detector and calorimeter for the direct detection of the two decay electrons. NEMO-3 measures 10~kg of isotopes in the form of very thin radiopure foils (30-60~mg/cm$^2$) with a total surface area of 20~m$^2$. 
The NEMO collaboration is developing the SuperNEMO detector~\cite{supernemoa,supernemob} which is the next generation of NEMO detectors. SuperNEMO will be able to measure 100~kg of $^{82}$Se with a sensitivity of $T_{1/2}(\beta\beta 0\nu)>10^{26}$~years which will be about two orders of magnitude higher than with NEMO-3. The isotope will be in the form of thin foils (40~mg/cm$^2$) that have a total surface area of 250~m$^2$. 

One of the troublesome sources of background for SuperNEMO is a possible contamination inside the source foils of $^{208}$Tl (Q$_{\beta}$~=~4.99~MeV) and $^{214}$Bi (Q$_{\beta}$~=~3.27~MeV) produced from the decay chains of $^{232}$Th and $^{238}$U respectively. 
In order to achieve the desired SuperNEMO sensitivity, the required radiopurities of the SuperNEMO double beta decay foils are $\mathcal{A}$($^{208}$Tl)~$<$~2~$\mu$Bq/kg and $\mathcal{A}$($^{214}$Bi)~$<$~10~$\mu$Bq/kg. 
A first radiopurity measurement of purified selenium samples can be performed by $\gamma$ spectrometry with ultra low background HPGe (High Purity Germanium) detectors. Nevertheless, the best detection limit that can be reached with this technique for $^{208}$Tl is around 50~$\mu$Bq/kg, which is one order of magnitude less sensitive that the required value. 
In order to achieve the required  sensitivity for SuperNEMO, a R$\&$D program has been performed to develop the BiPo detector dedicated to the measurement of  ultra-low levels of 
contamination in $^{208}$Tl and $^{214}$Bi in the foils of SuperNEMO.  
The final BiPo detector must be able to measure 12~m$^2$ of SuperNEMO double beta decay foils, corresponding to one SuperNEMO module. The measurement must not exceed a few months. Moreover, it must be able to qualify the radiopurity of the foils in their final form,  before installation into the SuperNEMO detector.
The BiPo detector also measures the surface radiopurity of other  materials.

A BiPo prototype, called BiPo-1, with an active surface of 0.8~m$^2$, has been built to validate the measurement technique, and to measure the different components of the background. BiPo-1  has been running since May 2008 in the Modane Underground Laboratory.
The first phase of data collection with BiPo-1 ran until June 2009 and was dedicated to the background measurement of $^{212}$Bi ($^{208}$Tl). 
In a second phase, new electronics was installed in July 2009 in order to measure simultaneously the $^{212}$Bi and $^{214}$Bi background. In this article, the results of $^{212}$Bi ($^{208}$Tl) measurement from the data taken only during the first phase, are presented. 

Two other prototypes have also been built in order to explore different techniques for the final BiPo detector. These detectors are the BiPo-Phoswich prototype which uses the phoswich technique~\cite{phoswich1,phoswich2} and the BiPo-2 prototype which uses a more compact geometry~\cite{bipo2}. The results obtained with these two prototypes will be given in a future publication. 

The experimental principle of the BiPo detector and the possible components of observable backgrounds are described in sections~\ref{sec:principle} and \ref{sec:backgrounds}. The description of BiPo-1 and the detection efficiency calculated by simulation are given in sections~\ref{sec:bipo1prototype} and \ref{sec:detectionefficiency}. The experimental validation of the BiPo technique, the results of the background measurements in the BiPo-1 detector and the extrapolated sensitivity to a larger 12~m$^2$ BiPo detector are presented in section~\ref{sec:bipo1results}.

\section{Measurement principle of the BiPo detector}
\label{sec:principle}

\begin{figure}[htb]
  \centering
  \includegraphics[scale=0.35]{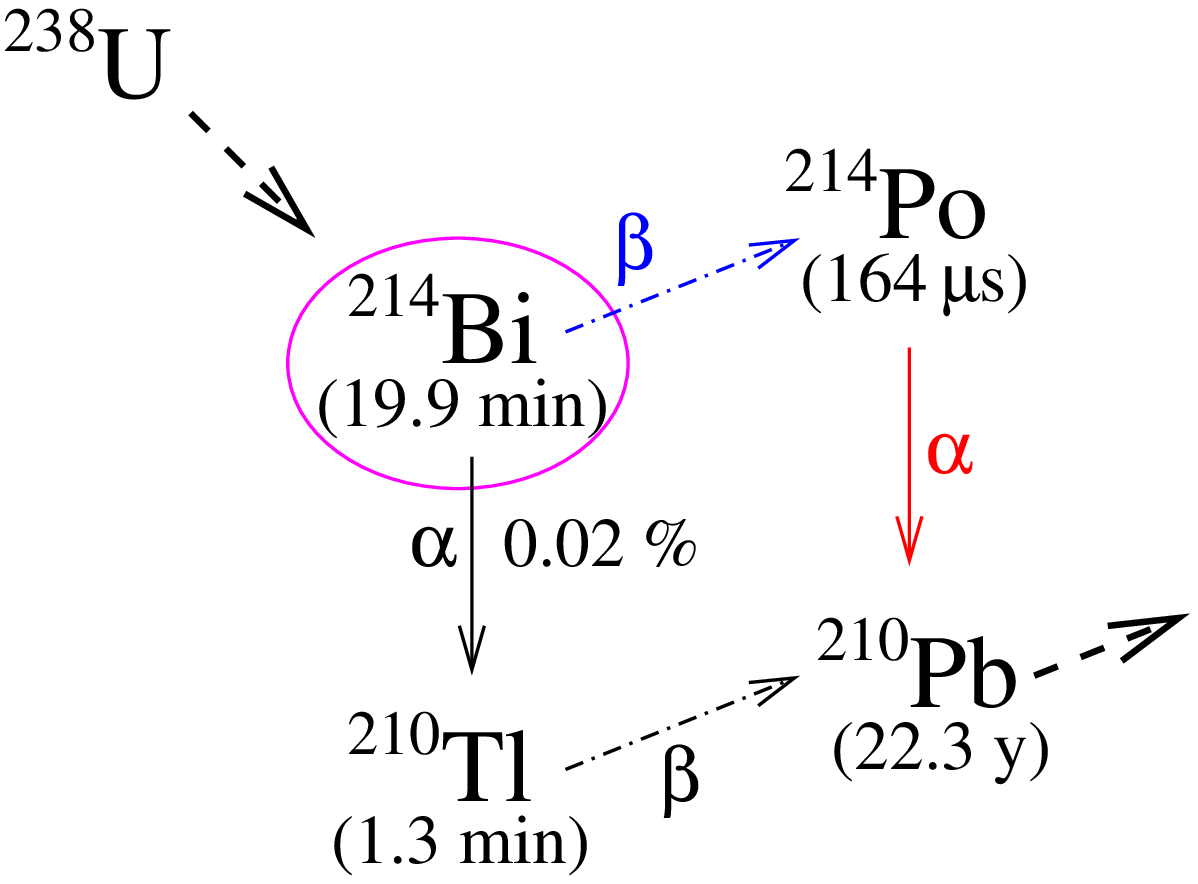}
  \hspace{1.5cm}
  \includegraphics[scale=0.35]{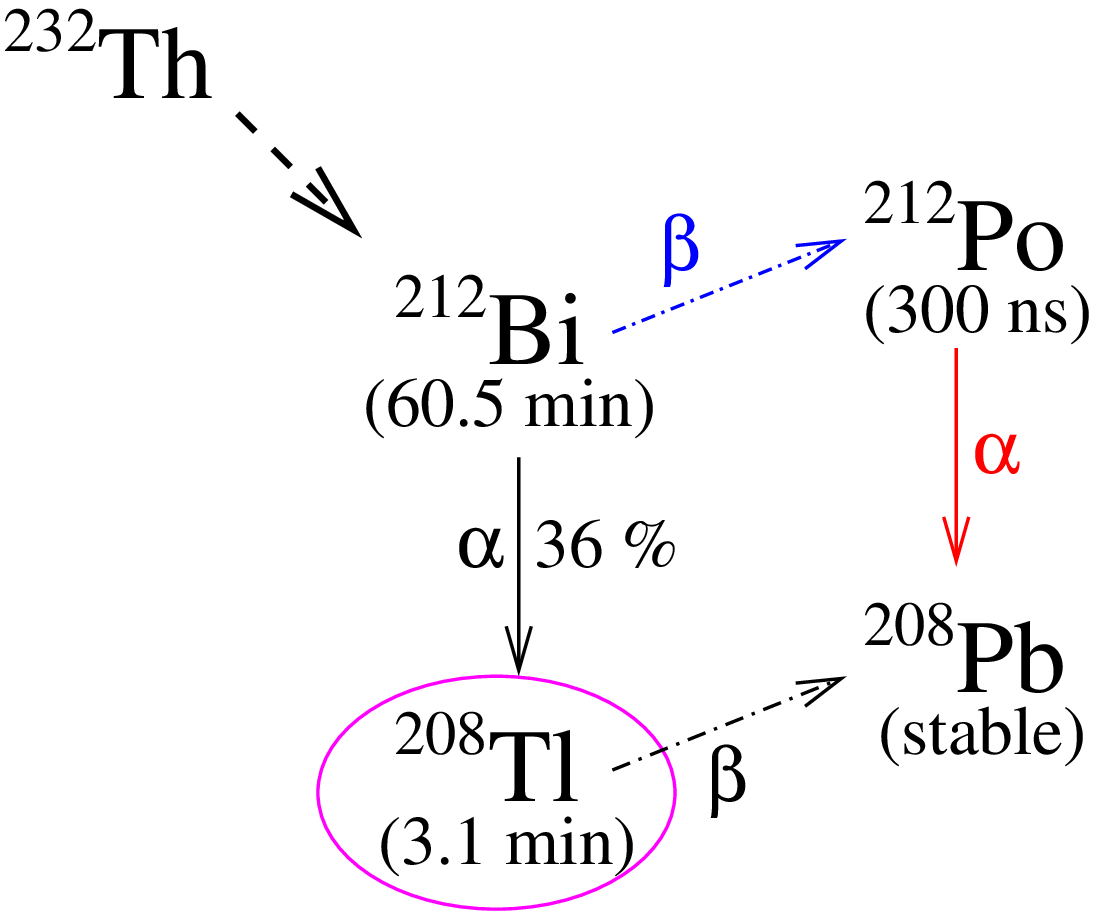}
  \caption{BiPo processes for $^{214}$Bi and $^{208}$Tl measurement.}
  \label{fig:bipo-proc}
\end{figure}

In order to measure $^{208}$Tl and $^{214}$Bi contaminations, the underlying concept of the BiPo detector is to detect with organic plastic scintillators the so-called BiPo process, which corresponds to the detection of an electron followed by a delayed $\alpha$ particle. The $^{214}$Bi isotope is a ($\beta$,$\gamma$) emitter (Q$_{\beta}$~=~3.27~MeV) decaying to $^{214}$Po, which is an $\alpha$ emitter with a half-life of 164~$\mu$s. The $^{208}$Tl isotope is measured by detecting its parent,  $^{212}$Bi. Here $^{212}$Bi decays with a branching ratio of 64\% via a $\beta$ emission (Q$_{\beta}$~=~2.25~MeV) towards the daughter nucleus $^{212}$Po which is a pure $\alpha$ emitter (8.78~MeV) with a short half-life of 300~ns (Figure~\ref{fig:bipo-proc}).

\begin{figure}[htb]
  \centering
  \includegraphics[scale=0.55]{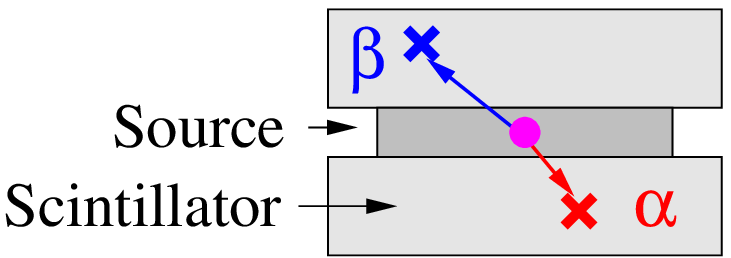}
  \hspace{1.5cm}
  \includegraphics[scale=0.35]{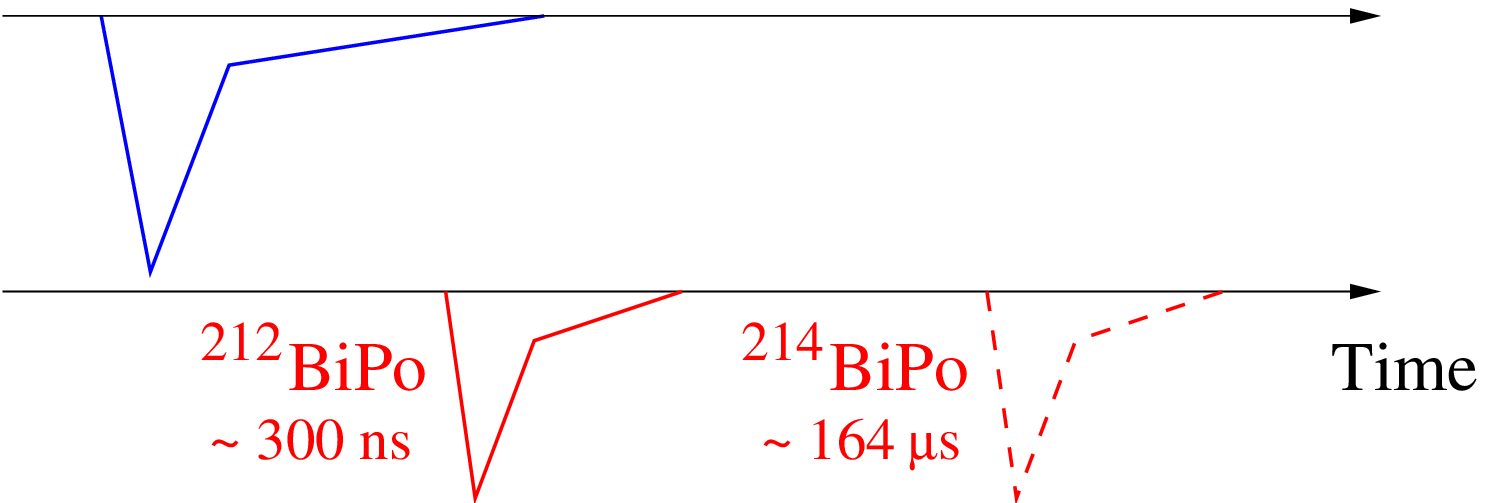}
  \caption{The BiPo detection principle with plastic scintillators and the time signals seen with PMTs for a {\it back-to-back} BiPo event. On the left image, the red dot represents the contamination and the crosses represent energy depositions in the scintillators (prompt signal in blue and delayed signal in red).}
  \label{fig:bipo-event}
\end{figure}

The BiPo experimental intent is to install the double beta decay source foil of interest between two thin ultra-radiopure organic plastic scintillators. The $^{212}$Bi ($^{208}$Tl) and $^{214}$Bi contaminations inside the foil are then measured by detecting the $\beta$ decay as an energy deposition in one scintillator without a coincidence from the opposite side, and the delayed $\alpha$ as a delayed signal in the second opposite scintillator without a coincidence in the first one. Such a BiPo event is identified as a {\it back-to-back} event since the $\beta$ and $\alpha$ enter different scintillators on opposite sides of the foil. The timing of the delayed $\alpha$ depends on the isotope to be measured (Figure~\ref{fig:bipo-event}).
The energy of the delayed $\alpha$ provides information on whether the contamination is on the surface or in the bulk of the foil.

A second topology of BiPo events can in principle be used. This involves the {\it same-side} BiPo events for which the prompt $\beta$ signal and the delayed $\alpha$ signal are detected in the same scintillator without a coincidence signal in the scintillator on of the opposite side. The detection of the {\it same-side} events would increase by about 50\% the BiPo efficiency. However, it will be shown in section~\ref{sec:surf-bkg} that the level of background measured in BiPo-1 using the {\it same-side} topology is much larger than the one measured in the {\it back-to-back} topology. For this reason, only {\it back-to-back} topology will be used for the rest of the BiPo-1 analysis presented in this paper.

\section{The different components of the background}
\label{sec:backgrounds}

\subsection{Random coincidences}
\label{subsec:randomcoincidences}

The first limitation of the BiPo detector is the rate of random coincidences between the two scintillators giving a signal within the delay time window (Figure~\ref{fig:bipo-background}a).
A delay time window equal to about three times the half-life of the  $\alpha$ decay is choosen (1~$\mu$s for $^{212}$Bi and 500~$\mu$s for $^{214}$Bi) in order to contain a large part of the signal and to minimize the random coincidence.
The single counting rate is dominated by Compton electrons due to external $\gamma$. BiPo-1 were then built with thin scintillators, with low radioactivity materials and installed inside a low radioactivity shield in an underground laboratory in order to reduce the $\gamma$ event rate. Additionally pulse shape analysis of the delayed signal is performed in order to discriminate between  electrons and  $\alpha$ events, thus rejecting random Compton electron coincidences due to external $\gamma$.

\subsection{Radiopurity of the scintillators}
\label{subsec:scintradiopurity}

\begin{figure}[htb]
  \centering
  \includegraphics[scale=0.6]{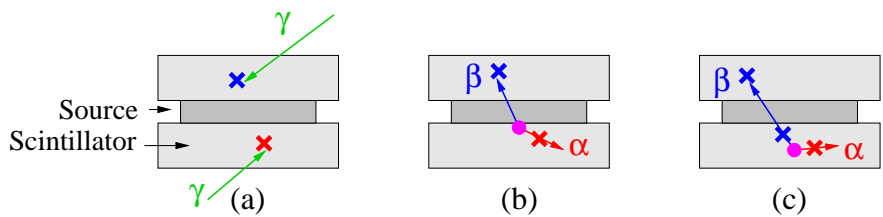}
  \caption{Illustration of the possible sources of background: (a) random coincidences due to the $\gamma$ flux, (b) bismuth contamination on the surface of the scintillators and (c) bismuth contamination in the volume of the scintillator.}
  \label{fig:bipo-background}
\end{figure}

The second source of background that mimics a BiPo event comes from bismuth contamination on the surface of the scintillator that is in contact with the foil. This surface contamination of the scintillators produces a signal indistinguishable from the true BiPo signal coming from the source foil, as shown in Figure~\ref{fig:bipo-background}b. 

In principle, bismuth contamination in the scintillator volume (bulk contamination) is not a source of background, because the emitted electron should trigger one scintillator block before escaping and entering the second one, as shown in Figure~\ref{fig:bipo-background}c. The two fired scintillator blocks are in coincidence and this background event is rejected.
However, if the contamination is not deep enough inside the scintillator but quite near the surface, the electron from the $^{212}$Bi-$\beta$ decay will escape the first scintillator and will fire the second one without depositing enough energy to trigger the first one. It will appear exactly like a BiPo event emitted from the foil.

\subsection{Radon and Thoron}
\label{subsec:radonthoron}

Other possible sources of background are via thoron ($^{220}$Rn) or radon ($^{222}$Rn) contamination of the gas between the foil and the scintillators. Thoron and radon decay to $^{212}$Bi and $^{214}$Bi respectively and a bismuth contamination on the surface of the scintillators is thus observed.
In order to suppress this source of background, radiopure gas and materials without thoron and radon degazing around the scintillators are required.

\section{The BiPo-1 prototype}
\label{sec:bipo1prototype}

\subsection{Description of the detector}
\label{subsec:description}

\begin{figure}[htb]
  \centering
  \includegraphics[scale=0.45]{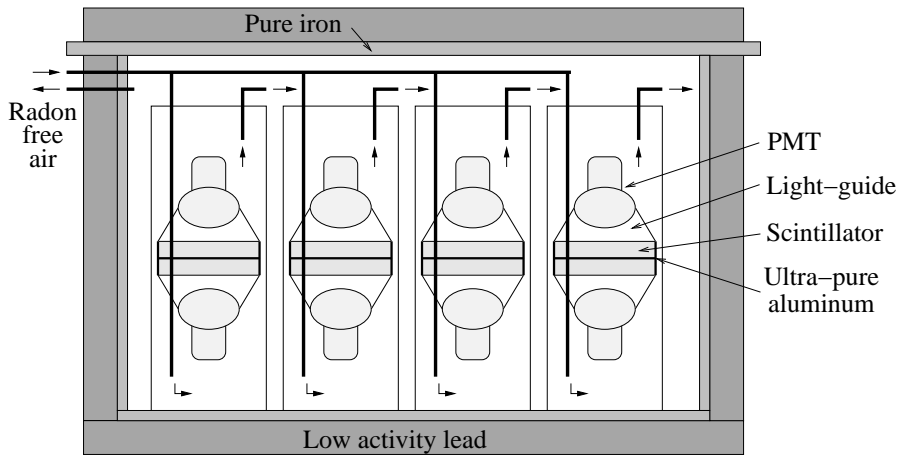}
  \caption{Schematic view of the BiPo-1 prototype inside its shield.}
  \label{fig:bipo1-setup}
\end{figure}

The BiPo-1 detector is composed of 20 similar modules. Each module consists of a gas and light tight box, containing two thin polystyrene-based scintillator plates of dimension 200$\times$200$\times$3~mm$^3$ which are placed face-to-face. Each scintillator plate is coupled to a low radioactivity 5'' photomultiplier (PMT R6594-MOD from Hamamatsu) by a UV Polymethyl Methacrylate light guide (Figure~\ref{fig:bipo1-setup}). Each module has a sensitive surface area of 0.04~m$^2$.
The scintillators have been prepared with a mono-diamond tool from scintillator blocks produced by JINR (Dubna, Russia) for NEMO-3. 
The surface of the scintillators facing the source foil has been covered with 200~nm of evaporated ultra-pure aluminum in order to optically isolate each scintillator and to improve the light collection efficiency. 
The entrance surface of the scintillators has been carefully cleaned before and after aluminium deposition using the  following cleaning sequence: acetic acid, ultra pure water bath, di-propanol and finally a second ultra pure water bath. 
The sides of the scintillators and light guides are covered with a 0.2~mm thick Teflon layer for light diffusion. 
All the materials of the detector have been selected by HPGe measurements to confirm their high radiopurity. 
Table~\ref{tab:radioactivities} gives a summary of the radioactivity HPGe measurements for the BiPo-1 materials. 
The modules are shielded by 15~cm of low activity lead which addresses the external $\gamma$ flux. The upper part of the shield which supports the lead is a pure iron plate 3~cm thick. 
Radon-free air ($\mathcal{A}$(radon)~$<$~1~mBq/m$^3$) flushes the volume of each module and also the inner volume of the shield. 

The first three modules of BiPo-1 were initially installed in the new Canfranc Underground Laboratory in Spain. Due to the temporary closure of this Laboratory, BiPo-1 was completed in the Modane Underground Laboratory where it has been running since May 2008.

\begin{table}[htb]
\centering
\begin{tabular}{c|c|c|c|c}
Type of materials  & $^{232}$Th & $^{238}$U & $^{40}$K & $^{210}$Pb    \\
\hline
\hline
Photomultipliers   & 0.028      & 0.48      & 1.0      &    \\
\hline
Module containers  & 0.09       & 0.06      & 0.7      &    \\
(carbon fiber)     &            &           &          &    \\
\hline
Black polyethylene &            & 0.01      & 0.2      &    \\
\hline
Screws             &            &           &          & 1.1  \\
\end{tabular}
\caption{Radioactivity HPGe measurements for the BiPo-1 materials. Activities are given in Bq per BiPo-1 module. Only materials with significant activities are presented.}
\label{tab:radioactivities}
\end{table}

Photomultiplier signals are sampled with MATACQ VME digitizer boards~\cite{breton}, during a 2.5~$\mu$s window with a high sampling rate (1~GS/s), 12-bit resolution and a dynamic range (1~volt). The level of electronic noise is $\sigma$=250~$\mu$V. The single photoelectron level is 1~mV.
The acquisition is triggered each time a PMT pulse reaches 50~mV (corresponding to 100~keV).

\subsection{Energy and time calibration}

  The linearity and the gain stability of all PMTs were initially measured on a PMT test bench originally developed for the LHCb detector~\cite{lhcb}.
During the assembly, each BiPo-1 module was also tested and calibrated in energy. 
An $^{241}$Am source emitting 5.5~MeV $\alpha$ particles and a $^{207}$Bi source emitting 1~MeV conversion electrons have been used.
The test bench was sucessfully used to measure the response of the BiPo-1 modules over the entire surface of the scintillators.  A decrease of 30$\%$ in the light collection was observed when the electron or $\alpha$ source was moved from the center of the scintillator to the edges. This non-uniformity is due to the fact that the PMT has been positioned very close to the scintillator (6~mm) in order to maximize the amount of detected light and to achieve an energy threshold as low as possible. 
This non-uniformity of the light collection produces an uncertainty of about 30$\%$ on the value of the energy threshold. This uncertainty is negligible for the BiPo detection efficiency as it will be shown later in section~\ref{sec:detectionefficiency} and Figure~\ref{fig:bipo-efficiency}. Moreover the energy measurement is not the critical issue for BiPo-1 since a BiPo event is recognized by its delay time and topology. 

\begin{figure}[htb]
  \centering
  \begin{minipage}[c]{0.3\linewidth}
  \includegraphics[scale=0.65]{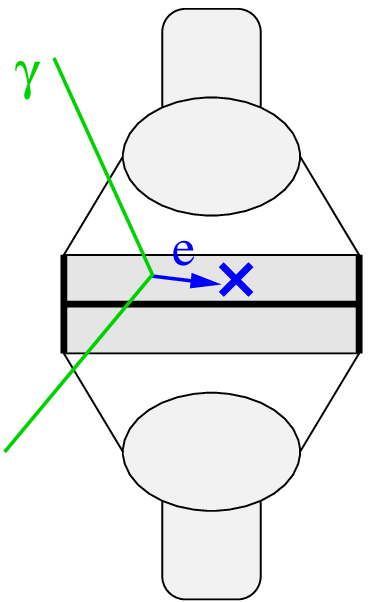}
  \end{minipage}
  \hfill
  \begin{minipage}[c]{0.65\linewidth}
  \includegraphics[scale=0.3, angle=270]{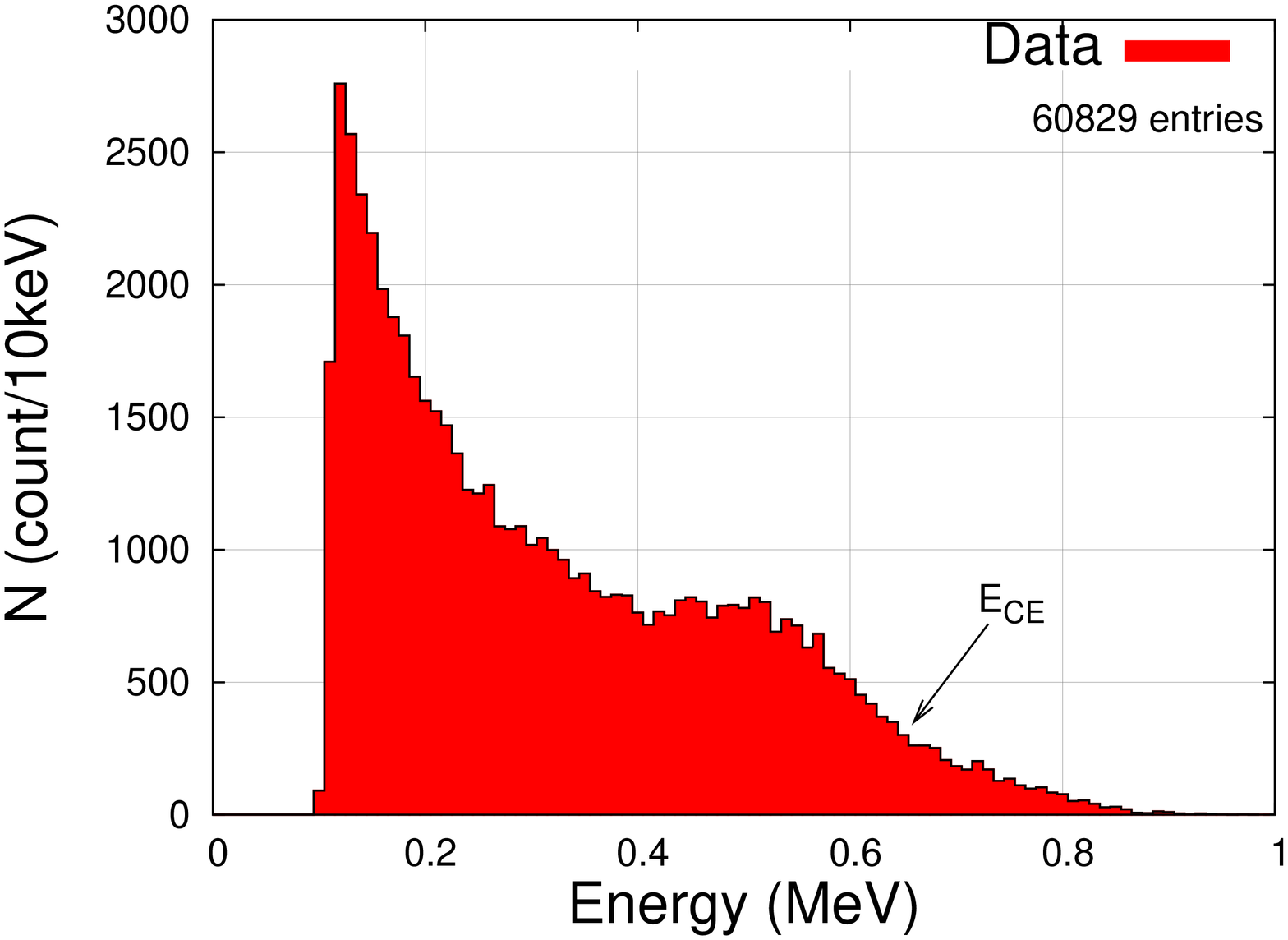}
  \end{minipage}
  \caption{Energy calibration with a $^{54}$Mn source: (left) showing the principle of the method and (right) an example of the energy spectrum obtained with a scintillator.}
  \label{fig:bipo1-calib-ener}
\end{figure}

\begin{figure}[htb]
  \centering
  \begin{minipage}[c]{0.3\linewidth}
  \includegraphics[scale=0.65]{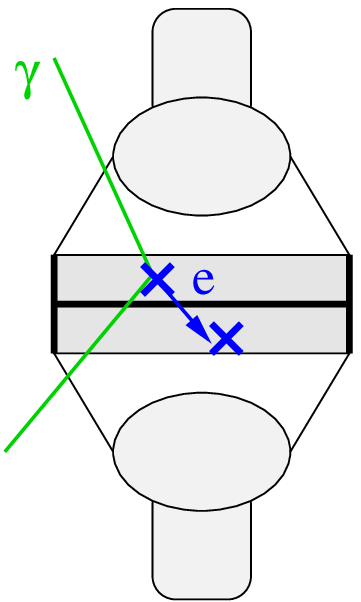}
  \end{minipage}
  \hfill
  \begin{minipage}[c]{0.65\linewidth}
  \includegraphics[scale=0.3, angle=270]{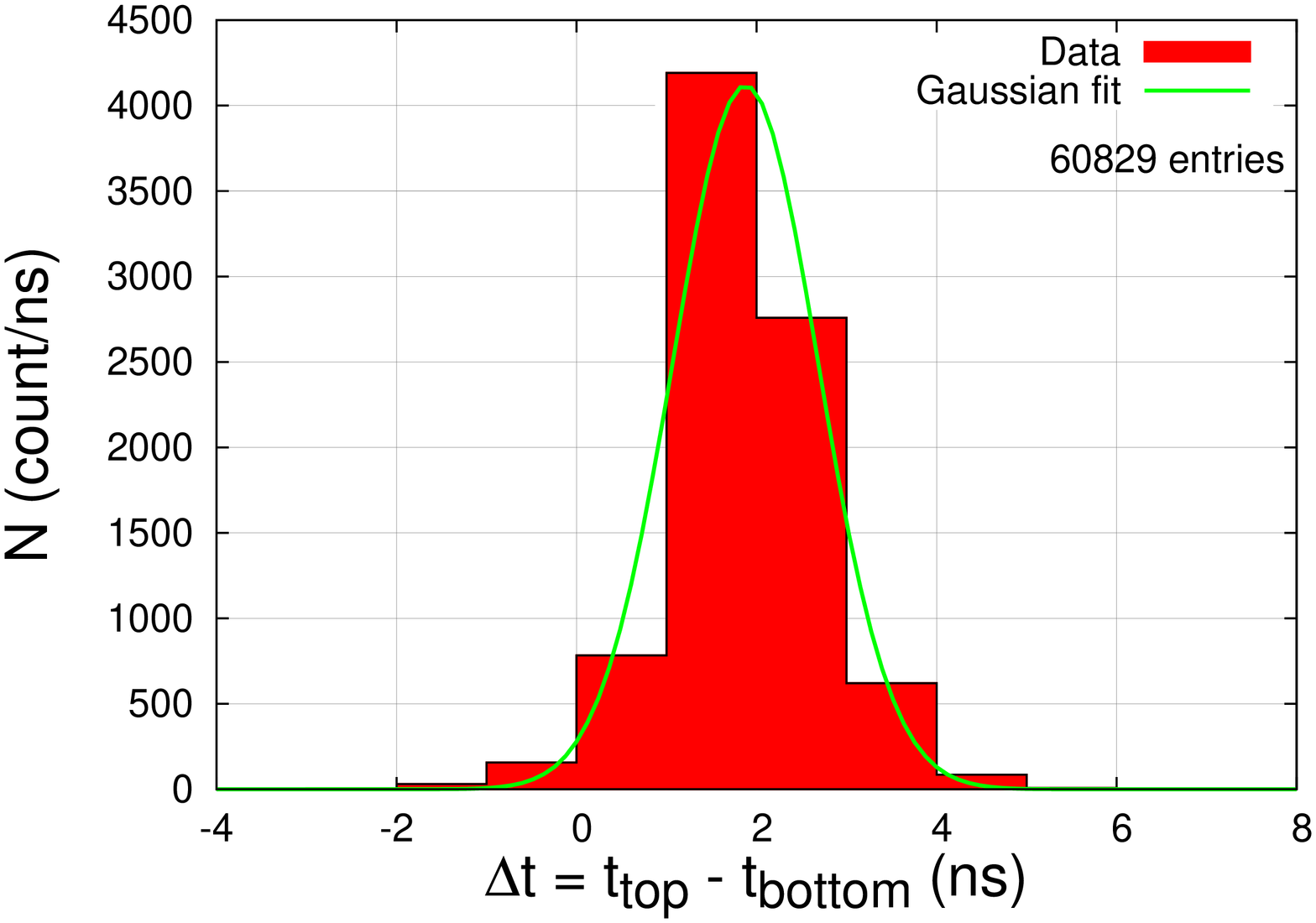}
  \end{minipage}
  \caption{Time calibration with a $^{54}$Mn source: (left) showing the principle of the method and (right) an example of the time distribution obtained with one module.}
  \label{fig:bipo1-calib-time}
\end{figure}

The energy and time calibrations of the 20-module BiPo-1 detector with its low radioactivity shield, were performed using a $^{54}$Mn source ($\gamma$ of 835~keV). The energy calibration was performed by measuring the Compton edge at 639~keV for electrons fully contained in one scintillator (Figure~\ref{fig:bipo1-calib-ener}).  The time calibration is performed by using a Compton electron produced in one scintillator which crossed over into the second scintillator (Figure~\ref{fig:bipo1-calib-time}). The average time difference between the two PMTs is measured and corrected in the  BiPo analysis software. 

\subsection{Selection of the BiPo events}

\begin{figure}[htb]
  \centering
  \includegraphics[scale=0.50]{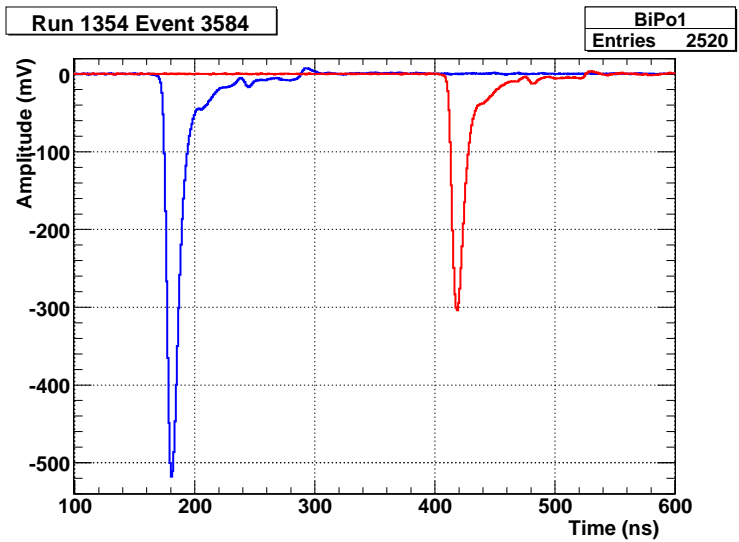}
  \caption{Example of a BiPo event observed in BiPo-1.}
  \label{fig:bipo1-event}
\end{figure}

The search for a BiPo cascade is performed in the software analysis by a study of the two sampled and recorded PMT signals from a given BiPo module. The {\it back-to-back} BiPo event is recognized as a prompt signal above 100~keV without any coincident signal above 10~keV in a second PMT of the opposite side. This signal is then followed by a delayed signal above 150~keV in the second PMT (delayed $\alpha$ particle) without a coincident signal above 10~keV in the first PMT.  
A minimum delay time of 20~ns is required in order to reject possible correlated signals due to an external particle shower produced by a cosmic ray or a high energy $\gamma$. The maximum delay time of 2.35~$\mu$s, given by the sampling board, corresponds to an efficiency of 99.6$\%$ to tag the delayed $\alpha$ from $^{212}$Po. A true BiPo event observed with BiPo-1 is shown in Figure~\ref{fig:bipo1-event}.

\section{Detection efficiency}
\label{sec:detectionefficiency}

\subsection{Energy quenching of $\alpha$ particles}
\label{subsec:energyquenchingofalphaparticles}

The efficiency  of the BiPo detector is limited by the capacity of an $\alpha$ to escape from the source foil being measured, and then deposit an amount of scintillation light above the detection threshold. 
Due to the very large stopping power for $\alpha$ particles and subsequent large excitation-to-scintillation energy partition, the amount of light produced by an $\alpha$ is smaller than that produced by an electron of the same energy.

\begin{figure}[htb]
  \centering
  \includegraphics[scale=0.4, angle=270]{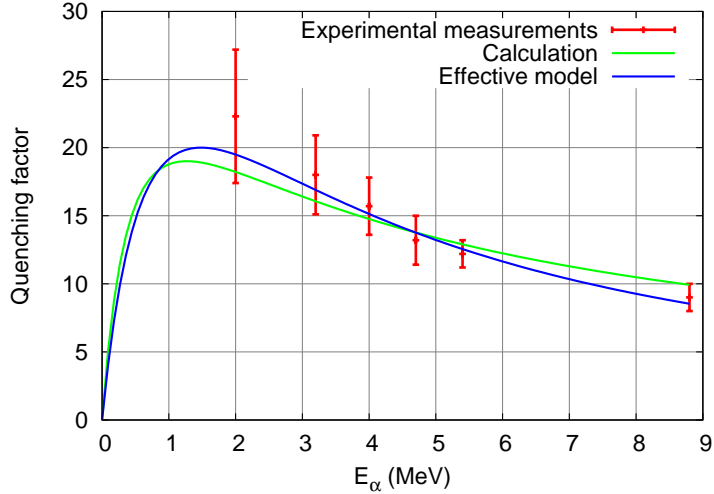}
  \caption{Quenching factors of $\alpha$'s in the organic plastic scintillator as a function of the energy deposited by the $\alpha$ on the scintillator: the red dots correspond to the measured values from this work, the green curve to the calculated value for polystyrene with a Birks factor $k_{B}=9.0 \times 10^{-3}$~g~MeV$^{-1}$~cm$^{-2}$ and the blue curve to the effective values used in the BiPo-1 simulations.}
  \label{fig:qf}
\end{figure}

A dedicated measurement of the quenching factor, defined as the ratio of the amount of light produced by an electron to the one produced by an alpha of the same energy, has been performed with the polystyrene-based plastic scintillators used in BiPo-1~\cite{bongrand} for different $\alpha$ energies. The $\alpha$ particles of 5.5~MeV emitted by an $^{241}$Am source were used. Their energies have been reduced by adding successively 6~$\mu$m Mylar foils between the source and the scintillator. A GEANT4 simulation of $\alpha$ particles emitted by $^{241}$Am and crossing several foils of Mylar has been done in order to determine the expected energy spectrum deposited in the scintillator. Then the quenching factor is calculated by comparing the observed energy with the energy obtained with 1~MeV electrons from a $^{207}$Bi source (Figure~\ref{fig:qf}).
The value of the quenching factor for $\alpha$'s with an energy of 8.78~MeV has been obtained separately using NEMO-3 data by analysing the BiPo decays on the surface of the NEMO-3 scintillators.

Recently a semi-empirical method of calculating the quenching factors for scintillators has been proposed and is described in~\cite{tretyak}. It is based on the Birks formula with the total stopping power for electrons and ions which are calculated with the ESTAR and SRIM codes. The method has only one free parameter, the Birks factor $k_{B}$~~\cite{birk}. The fit of the experimental data gives a Birks factor, $k_{B}=9.0 \times 10^{-3}$~g~MeV$^{-1}$~cm$^{-2}$, and is presented in Figure~\ref{fig:qf}. BiPo-1 measurements are in agreement with the calculation. 
An effective model of the quenching factor, which averages the measured and calculated values has been used for the BiPo-1 simulations and is presented in Figure~\ref{fig:qf}.

\subsection{Efficiencies}
\label{subsec:efficiencies}

The efficiency of the BiPo detector has been calculated with a GEANT4 Monte Carlo simulation. The $^{212}$BiPo events have been simulated assuming a uniform $^{212}$Bi contamination in the volume of a $^{82}$Se foil  40~mg/cm$^2$ (80~$\mu$m) thick. Only {\it back-to-back} BiPo events have been selected. An energy threshold  of 100~keV for the prompt signal and 150~keV for the delayed signal (corresponding to a threshold of about 2.5~MeV for the delayed $\alpha$), has been set.  
The BiPo events are rejected in the case of a back-scattering of the $\beta$ in one scintillator, with a deposited energy above the 10~keV threshold, before entering the second scintillator. Such events with two scintillators in prompt coincidence are rejected as candidates of $^{212}$Bi contamination in the scintillators (see section~\ref{sec:backgrounds}).
The total efficiency to detect a BiPo cascade is 5.5$\%$\footnote{If both the {\it back-to-back} and the {\it same-side} events are selected, the total efficiency becomes equal to 8.3\%}.

\begin{figure}[htb]
  \centering
  \includegraphics[scale=0.25, angle=270]{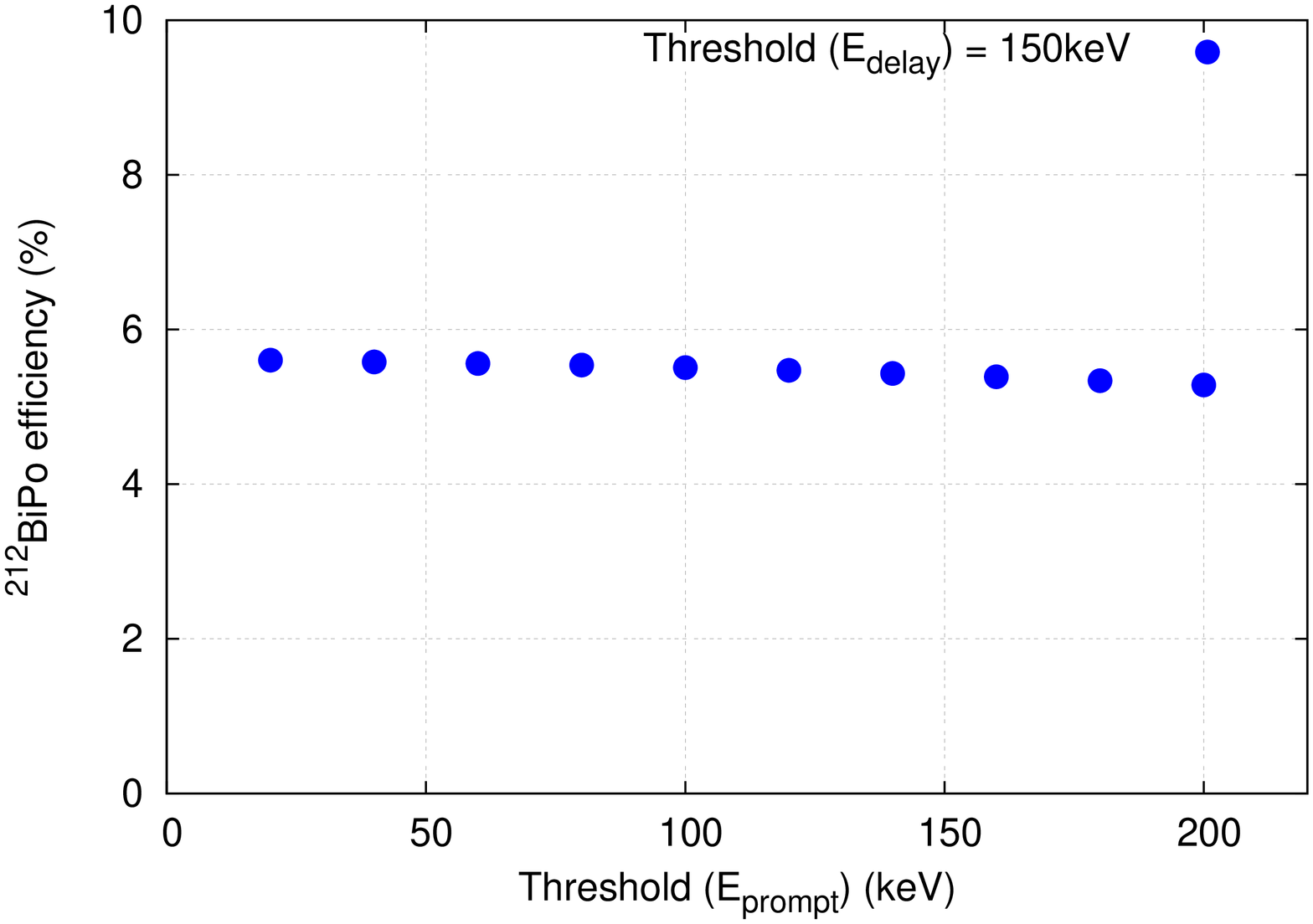}
  \hspace{0.0cm}
  \includegraphics[scale=0.25, angle=270]{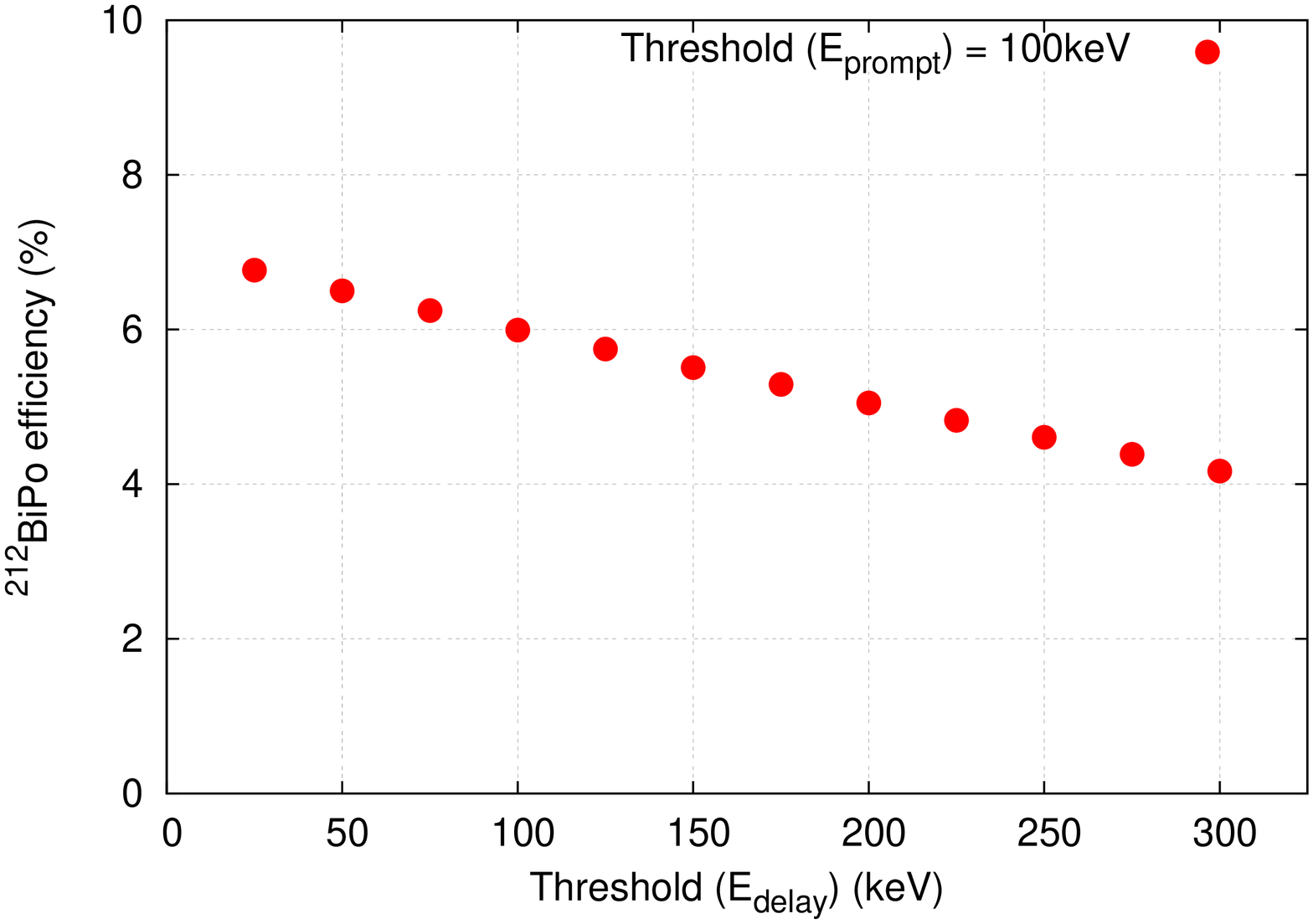}
  \caption{BiPo detection efficiency for $^{212}$Bi calculated by Monte Carlo simulations: (left) as a function of the energy threshold to detect the prompt signal assuming an energy threshold of the delayed signal of 150~keV; and (right) as a function of the energy threshold to detect the delayed signal assuming an energy threshold of 100~keV for the prompt signal.}
  \label{fig:bipo-efficiency}
\end{figure}

The systematic error on the efficiency is dominated by the uncertainty on the energy threshold for $\alpha$'s. The value of the energy threshold is correlated to three parameters: the exact values of the quenching factor of the $\alpha$'s, the BiPo-1 energy calibration,  and the effective corrections of the non-uniformity of the light collection along the surface of the scintillators. Taking into account the possible systematic errors on these three parameters, one estimates that the error on the energy threshold for $\alpha$'s is $\pm$100\%. 
The variation of the BiPo efficiency as a function of the energy threshold is shown in Figure~\ref{fig:bipo-efficiency}. The efficiency depends very slightly on the energy threshold of the prompt $\beta$ signal. However, the efficiency varies by $\pm$20\% if the energy threshold of the delayed $\alpha$ signal varies from 75~keV to 300~keV. Thus the systematic error of the BiPo efficiency is estimated to be 20\%.

The BiPo efficiency depends on the localization of the bismuth contamination in the foil. If the contamination is only on the surface of the foil, the BiPo efficiency is larger and has been calculated to be 10\%. 
The BiPo efficiency depends also on the composition of the foil. For instance with an aluminium foil 40~mg/cm$^2$ (150$\mu$m) thick which has been used to calibrate BiPo-1 (see section~\ref{subsec:principlevalidation}), the BiPo efficiency calulated by simulation is only 3.4$\%$. The difference in efficiency between aluminium and selenium foils is explained by their different Z/A ratios. This ratio is larger for aluminium (Z/A=13/27=0.48) than for $^{82}$Se (Z/A=34/82=0.41). Therefore the ionization by an $\alpha$ particle is larger in an aluminium foil, consequently, its probability to escape from it is smaller than from a selenium foil.
A summary of the calculated BiPo efficiencies with different cases is given in Table~\ref{tab:efficiency}. 

\begin{table}[htb]
\centering
\begin{tabular}{c|c|c}
Type of foil & Type of contamination & Efficiency \\
\hline
\hline
$^{82}$Se 40~mg/cm$^2$   & Bulk            & 5.5\% \\
\hline
$^{82}$Se 40~mg/cm$^2$   & Surface         & 10\% \\
\hline
Aluminium 40~mg/cm$^2$ & Bulk            & 3.4\% \\
\hline
No foil                  & Scintillators surface & 27\% \\
\end{tabular}
\caption{BiPo efficiencies calculated by Monte-Carlo simulations for different types of measurements.}
\label{tab:efficiency}
\end{table}

\section{Results from the BiPo-1 detector}
\label{sec:bipo1results}

\subsection{Experimental verification of the BiPo-1 technique with a calibrated aluminium foil}
\label{subsec:principlevalidation}

The first BiPo-1 module was dedicated to testing  the detection of bulk $^{212}$Bi contamination in a calibrated foil. A 150~$\mu$m thick aluminium foil (40~mg/cm$^2$) with an activity $\mathcal{A}$($^{212}$Bi$\rightarrow ^{212}$Po)~=~0.19~$\pm$~0.04~Bq/kg was first measured with low background HPGe detectors, and then installed between the two scintillators of the BiPo-1 module.

\begin{figure}[htb]
  \centering
  \includegraphics[scale=0.25, angle=270]{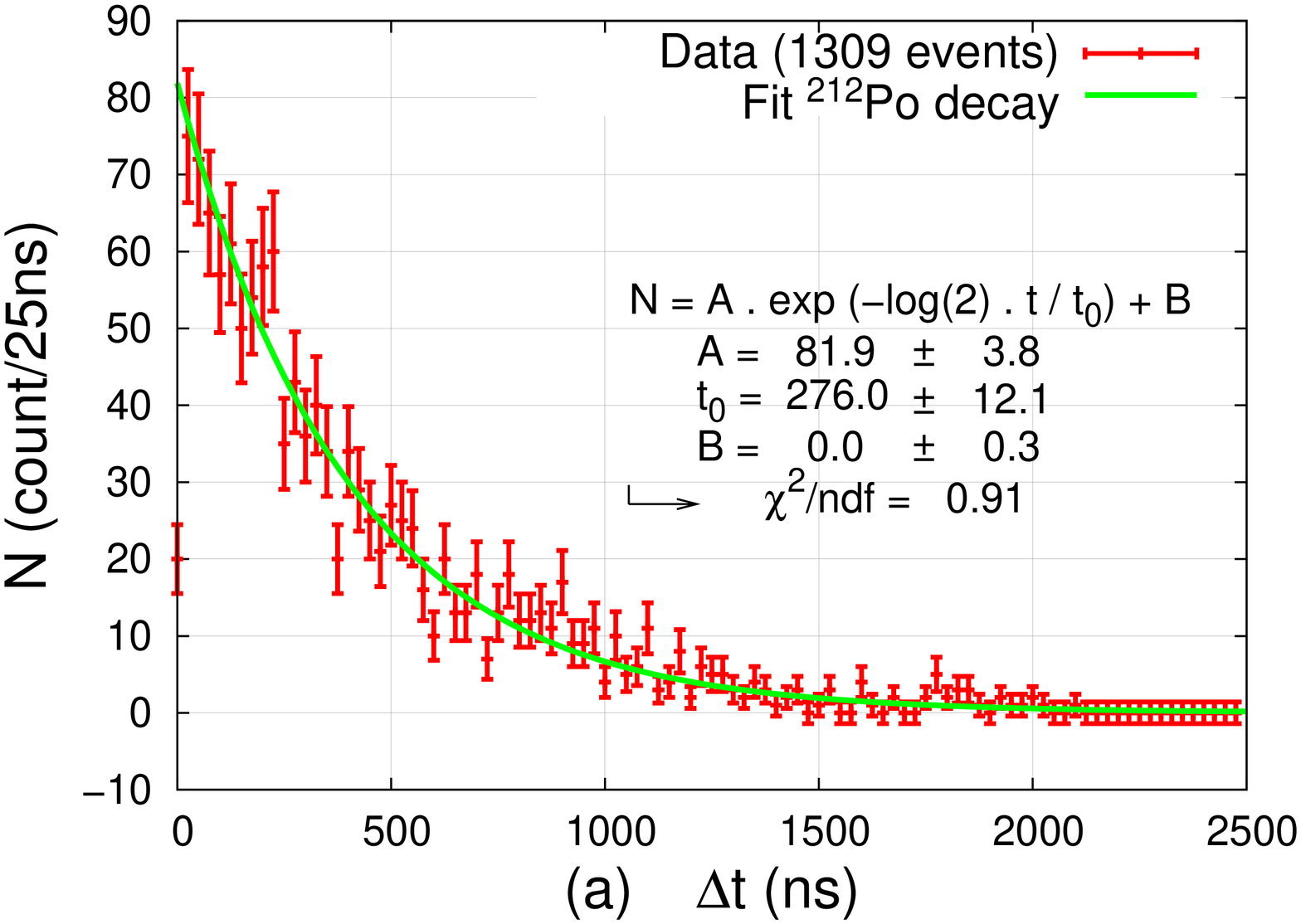}
  \includegraphics[scale=0.25, angle=270]{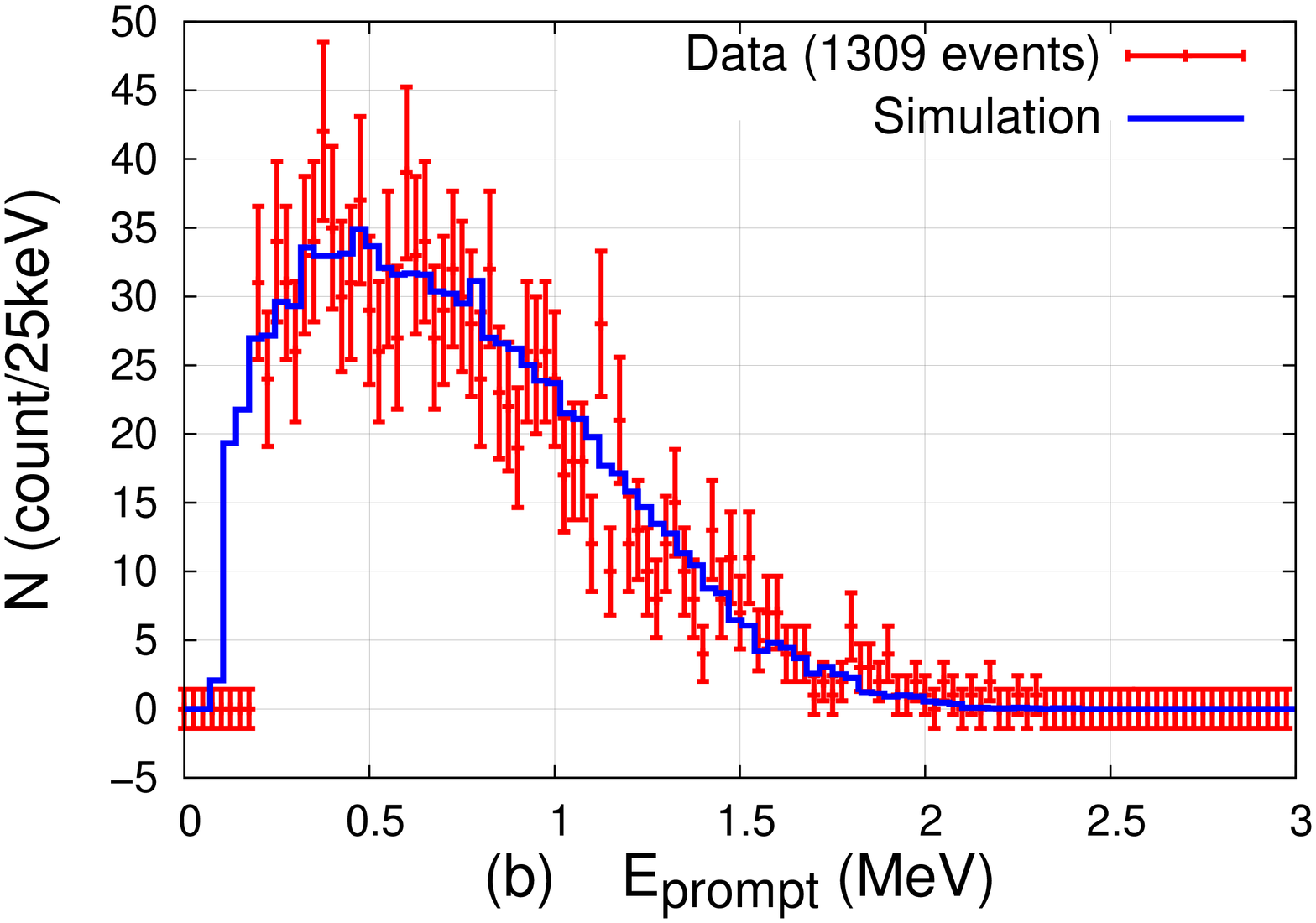}
  \includegraphics[scale=0.25, angle=270]{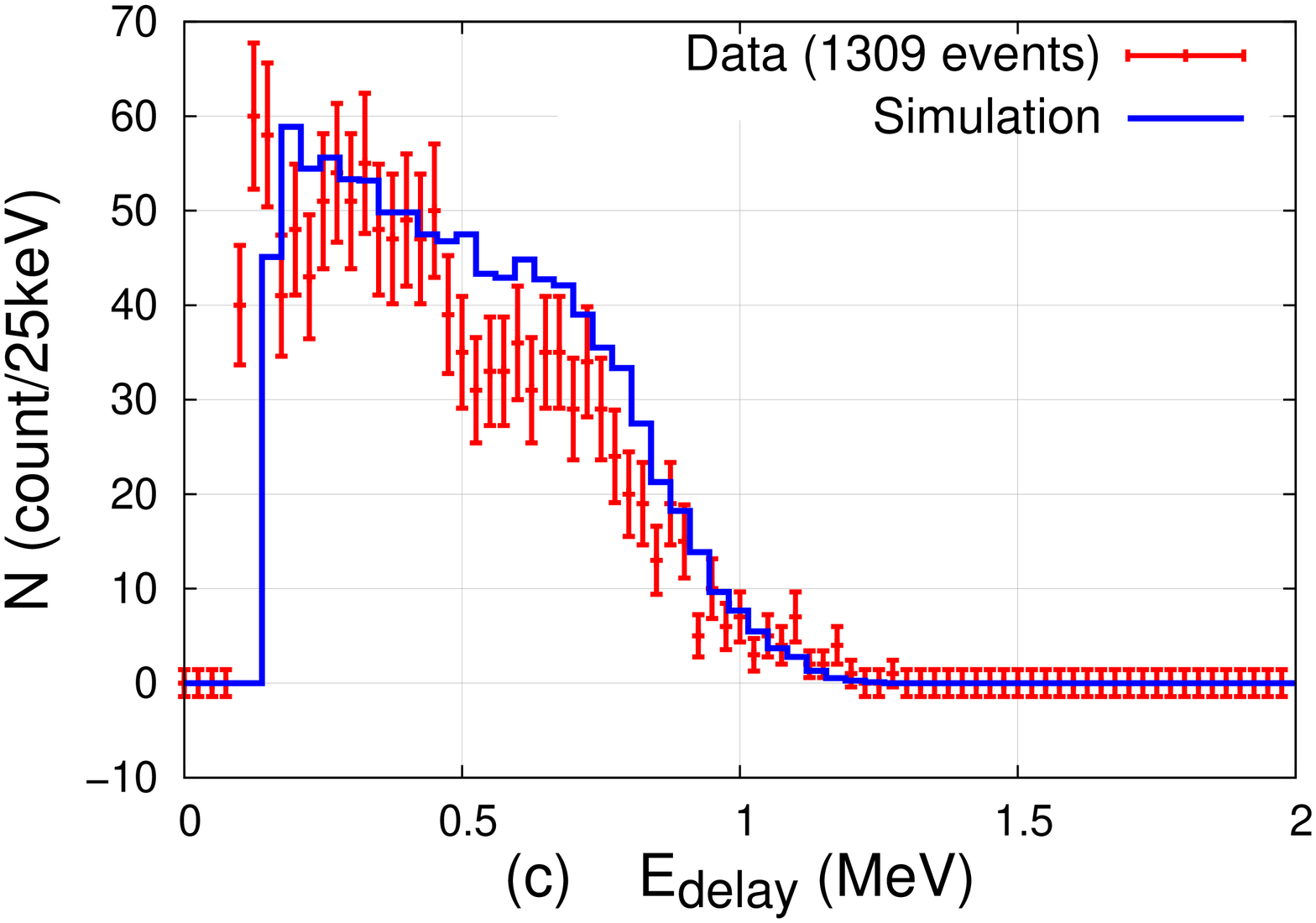}
  \caption{(a) The delay time distribution between the $\beta$ and $\alpha$ decays, (b) energy distribution of the prompt $\beta$ and (c) energy distribution of the delayed $\alpha$, for the 1309 BiPo events detected with the calibrated aluminium foil. The red dots are the data and the blue curve is the distribution calculated by simulations. The green curve in the delay time distribution is the results of the exponential decay fit.}
  \label{fig:bipo1-c1-results}
\end{figure}

After 160~days of data collection, a total of 1309 {\it back-to-back} BiPo events were detected. Taking into account the 3.4$\%$ efficiency calculated by GEANT4 simulations with a 20\% systematic error (see section~\ref{subsec:efficiencies}), this corresponds to an activity of $\mathcal{A}$($^{212}$Bi$\rightarrow ^{212}$Po)~$ = 0.16 \pm 0.005 (stat) \pm 0.03(syst)$~Bq/kg, in good agreement with the HPGe measurement. The distribution of the delay time between the two signals is presented in Figure~\ref{fig:bipo1-c1-results}. The half-life obtained from the fit is T$_{1/2}=276 \pm 12 (stat)$~ns. This measured half-life is in agreement with the experimental weighted average value of 299~ns for $^{212}$Po~\cite{po212}. These results confirm the measurement principle and the calculated efficiency. 

The energy spectra of the prompt $\beta$ and the delayed $\alpha$ are presented in Figure~\ref{fig:bipo1-c1-results}. There is good agreement with the expected spectra calculated by simulations. 
The  energy spectrum of the first  signal corresponds to a typical $\beta$ spectrum with $Q_{\beta}=2.25$~MeV. 
The energy spectrum of the delayed signal goes up to 1~MeV as expected for the $\alpha$ of 8.78~MeV from $^{212}$Po and a  quenching factor of Q$_f\sim$~9 at 8.78 MeV (see Figure~\ref{fig:qf}). No evidence of a peak at the 1~MeV endpoint in the energy spectrum of the delayed $\alpha$ indicates that the radioactive contamination is inside the volume and not on the surface of the aluminium foil.

\subsection{Discrimination of $\beta$ and $\alpha$ particles}
\label{subsec:eadiscrimination}

\begin{figure}[htb]
  \centering
  \includegraphics[scale=0.45]{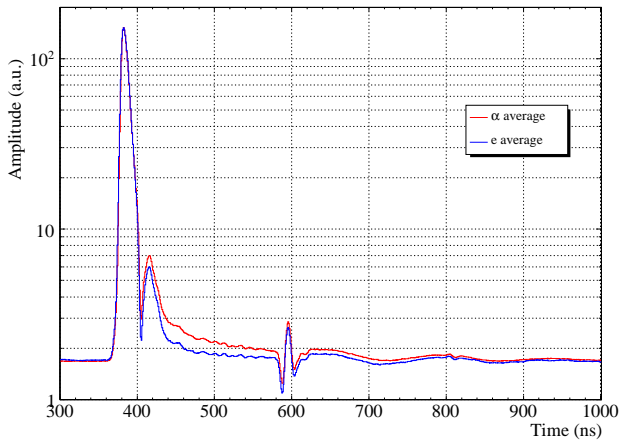}
  \caption{Average PMT signal obtained with a BiPo-1 module for 1 MeV electrons ($^{207}$Bi source) or 5.5 MeV $\alpha$  ($^{241}$Am). The amplitudes have been normalized to the first peak. The secondary peaks are due to electronic bounces in the PMT HV divider due to an imperfect impedance matching.}
  \label{fig:bipo1-discri-signal}
\end{figure}

Longer half-life states in the scintillators are excited by $\alpha$ particles but not by electrons due to the much larger energy loss for $\alpha$ particles. Thus, a higher tail of the signal produced by $\alpha$ particles is expected compared to electrons.  
The average PMT signal obtained with a BiPo-1 module for electrons ($^{207}$Bi source) or $\alpha$  ($^{241}$Am source) is presented in Figure~\ref{fig:bipo1-discri-signal}. A small but significant component in the tail of the signal is well observed for $\alpha$ particles compared to electrons. A pulse shape discrimination was developed using this set of data. The discrimination factor $\chi$ is defined as the ratio of the charge $q$ in the slow component to the total charge $Q$ of the signal. The charge $q$ is integrated from 15~ns after the signal peak to 900~ns. This integration window was optimized in order to maximize the discrimination. 

\begin{figure}[htb]
  \centering
  \includegraphics[scale=0.33]{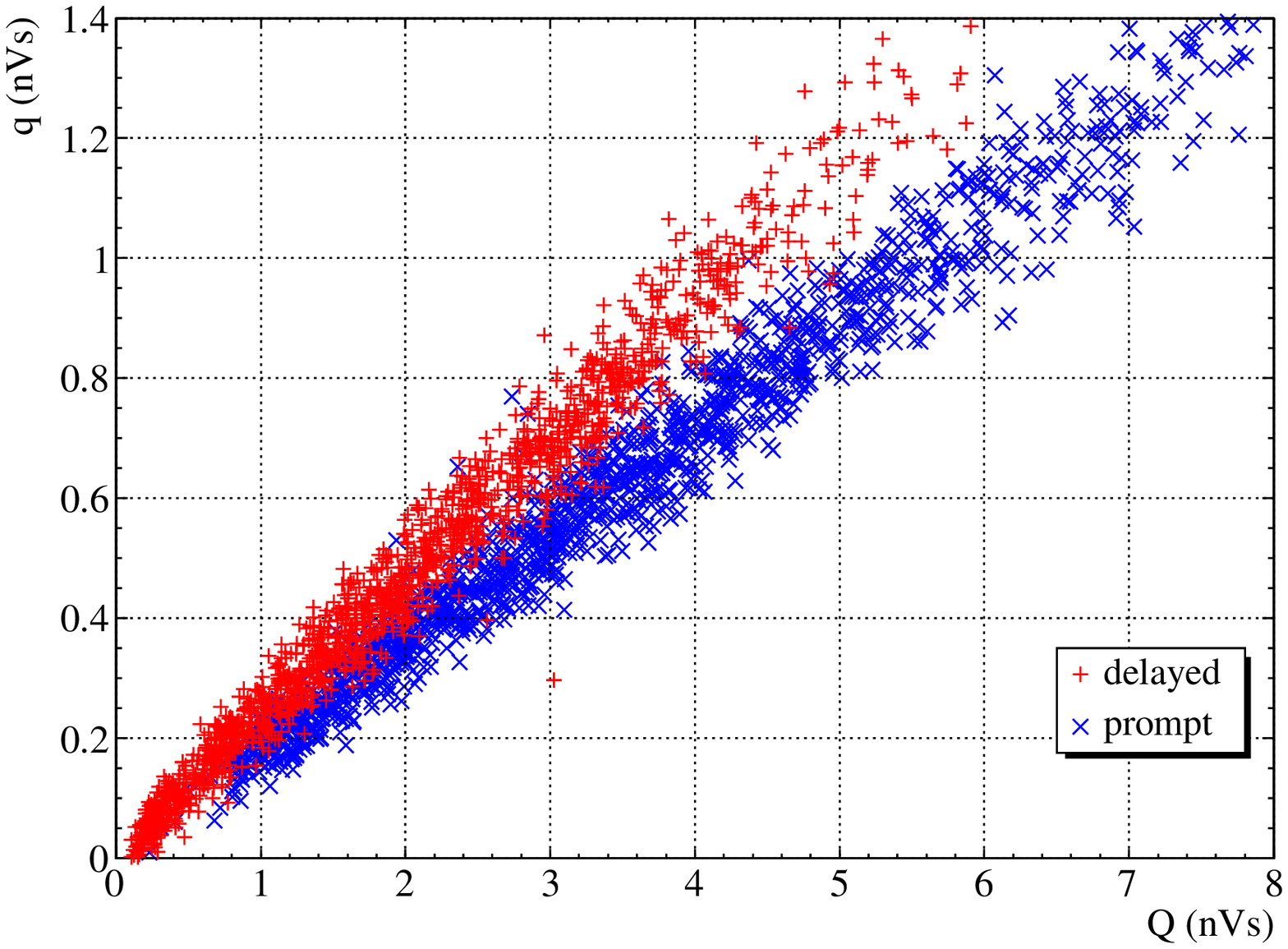}
  \includegraphics[scale=0.33]{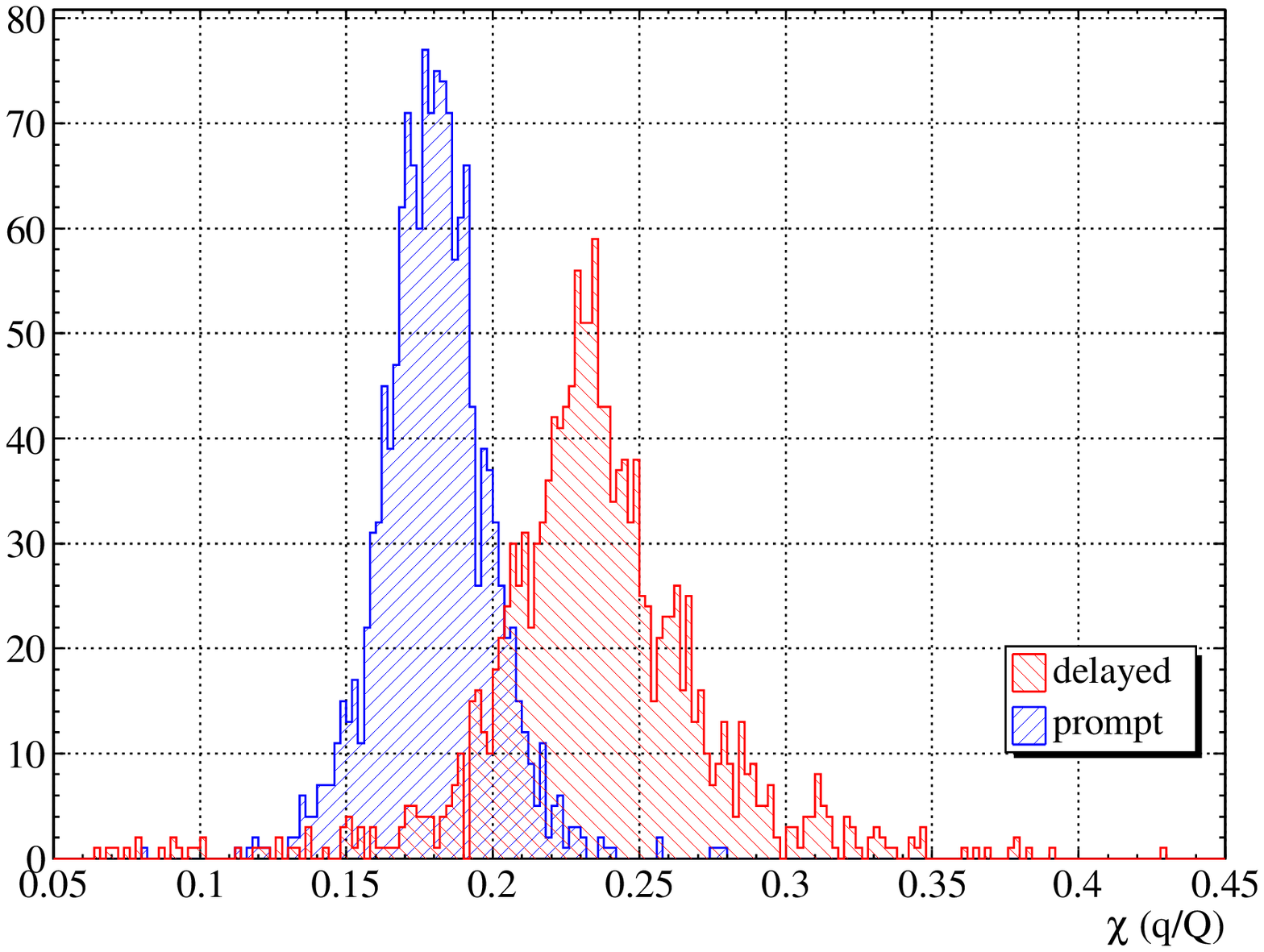}
  \caption{Discrimination of the prompt $\beta$  and delayed $\alpha$ signals using the BiPo events measured with the calibrated aluminium foil: (Left) Distribution of the charge $q$ in the slow component of the signal as a function of the total charge $Q$ of the signal; (Right) Distribution of the discrimination factor $\chi = {q \over Q}$.}
  \label{fig:bipo1-discri-caps1}
\end{figure}

\begin{figure}[htb]
  \centering
  \includegraphics[scale=0.45]{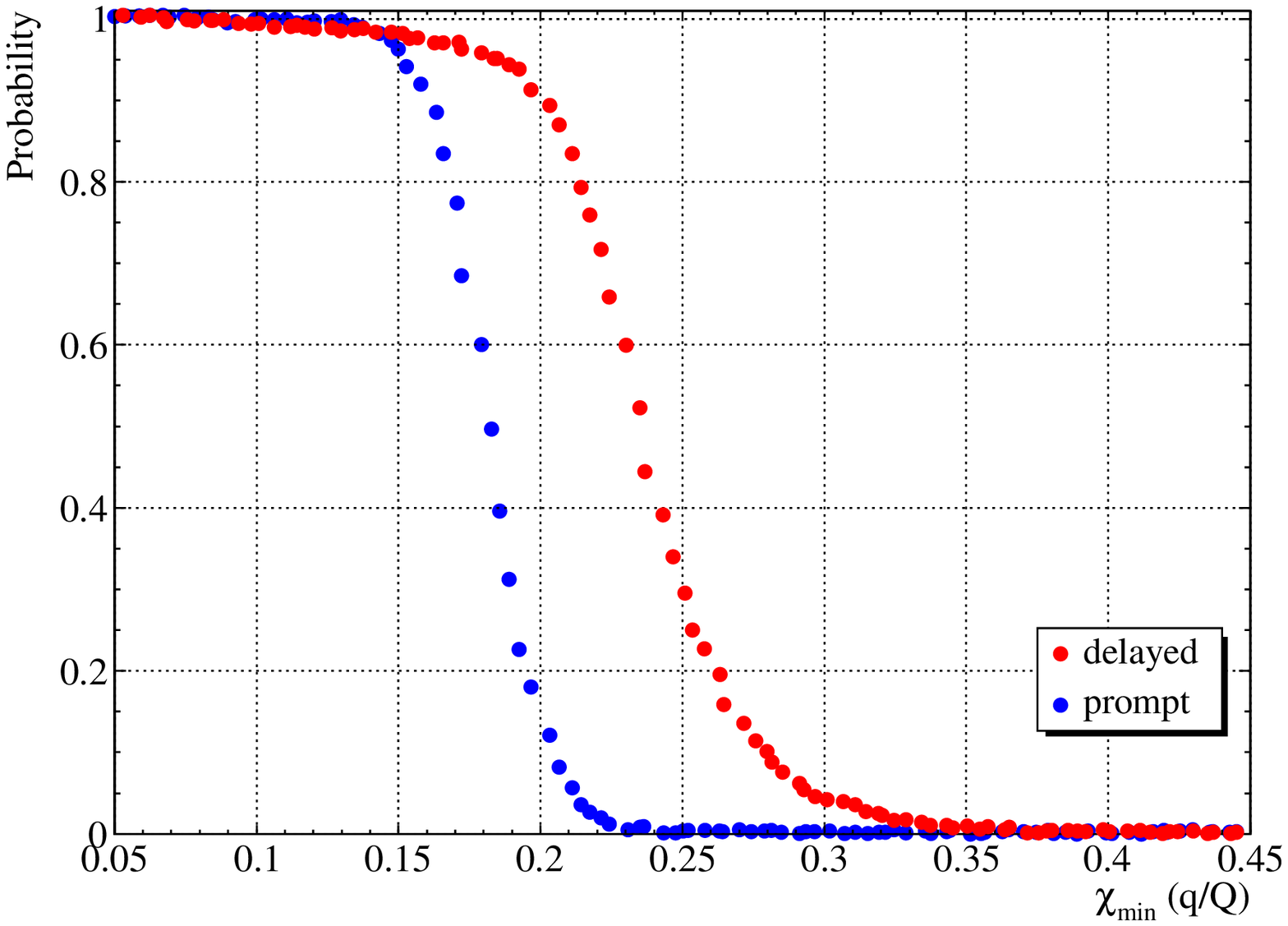}
  \caption{Probability of selecting an electron or an $\alpha$ as a function of the lower limit $\chi_{min}$ set by the discrimination factor $\chi$. The criterion $\chi > 0.2$ allows one to keep 90\% of the $\alpha$ particles and to reject 85\% of the electrons.}
  \label{fig:discri-prob-caps1}
\end{figure}

This electron and $\alpha$ pulse shape discrimination has been applied to the set of the BiPo events detected with the calibrated aluminium foil inside a first BiPo-1 module in order to calculate its global efficiency. As shown in Figure~\ref{fig:bipo1-discri-caps1}, a good separation is observed for prompt $\beta$ and delayed $\alpha$ signals although the discrimination becomes less efficient at low energy. By selecting a discrimination factor of $\chi > 0.2$ as applied to the delayed signal allows one to keep 90\% of the true BiPo events and reject 85\% of the random coincidences as is shown in Figure \ref{fig:discri-prob-caps1}. 

\subsection {Random coincidences}
\label{sec:random-coinc}

The single counting rate $\tau_0$ of a BiPo-1 scintillator without any coincidence in the opposite scintillator of the same module, was measured in the Modane Underground Laboratory with an energy threshold of 150 keV and $\tau_0 \approx 20$~mHz. This counting rate which is dominated by Compton electrons produced by external $\gamma$'s is stable and independent of the modules. It corresponds to an expected number of coincidences for the $^{212}$Bi measurement of $N_{coinc}$($^{212}$BiPo)~$= 2.10^{-3}$ events/(BiPo-1 module)/month, a negligible background. 

This result can be extrapolated to a larger BiPo detector with a sensitive surface of 12~m$^2$.
The number of random coincidences, $N_{coinc}$, is given by
$$N_{coinc} = 2 \,.\, N_{modules} \,.\, \tau_0^2  \,.\, \Delta T_{coinc} \,.\, T_{obs}$$
where $N_{modules}$ is the number of modules, $\Delta T_{coinc}$ is the delay time window and $T_{obs}$ is the duration of the measurement.
Given a detector composed of larger but thinner $300 \times 300 \times 1.5$~mm$^3$ scintillator plates, corresponding to a total of 130 modules, one can expect to have the same single counting rate $\tau_0$ since the volume of the scintillator plates is the same as the one in the BiPo-1 and.
The delay time window $\Delta T_{coinc}$ is chosen equal to three times the half-life of the delayed $\alpha$ decay of the polonium in order to contain $87.5\%$ of the BiPo events. 
The expected number of coincidences for the $^{212}$Bi measurement is $N_{coinc}$($^{212}$BiPo)~$= 0.25$~events/month, low enough to reach a sensitivity of 2~$\mu$Bq/kg in $^{208}$Tl as required by the SuperNEMO experiment. 

However, for the $^{214}$Bi measurement, the number of random coincidences $N_{coinc}$($^{214}$BiPo) becomes too large  because of the longer half-life of $^{214}$Po and $N_{coinc}$($^{214}$BiPo)~$= 135$~events/month.
This background can be reduced by applying the electrons-$\alpha$ discrimination defined earlier to the delayed signal in order to select only delayed $\alpha$ particles and reject the Compton electron coincidences. 
The number of random coincidences after electrons-$\alpha$ discrimination becomes $N_{coinc}$($^{214}$BiPo)~$= 20$~events/month. Given no other background component, this would correspond to a sensitivity of about 10~$\mu$Bq in $^{214}$Bi as required by the SuperNEMO experiment.

\subsection {Measurement of the scintillator bulk radiopurity}
\label{sec:bulk-bkg}

A BiPo-1 module has been dedicated to measure the bulk radiopurity of the organic polystyrene-based plastic scintillators produced by the JINR (Dubna, Russia) and used in BiPo-1. This module is similar to a standard BiPo-1 module except that it is equipped with two thicker scintillator blocks 20$\times$20$\times$10~cm$^3$ each, that are wrapped with aluminized Mylar. One of these blocks is from the NEMO-3 batch production used in BiPo-1. The second one is from a newer manufacturing process at JINR in 2007. 

The $^{212}$Bi contamination inside the scintillator blocks is recognized as a prompt signal from one PMT and a delayed signal of up to 1~$\mu$s from the same PMT. 
Since the delayed $\alpha$ is fully contained in the scintillator, its deposited energy in scintillation is expected to be around 1 MeV, due to the quenching factor. Thus, it is required that the energy of the delayed signal be greater than 700 keV (5 sigma less than 1 MeV). 
After 141 days of data collection, 10 events have been detected in the scintillator block from the NEMO-3 batch production and 24 events in the block from the new production. These numbers are in agreement with the expected number of BiPo events emitted from the aluminized Mylar surrounding the scintillator. However, if it is assumed that all the detected BiPo events come from a bulk contamination of the scintillators, a conservative upper limit on the contamination from $^{208}$Tl in the scintillators would be:

\begin{itemize}
\item NEMO-3 batch production used in BiPo-1: $\mathcal{A}$($^{208}$Tl)~$\leq$~0.13~$\mu$Bq/kg
\item New JINR production: $\mathcal{A}$($^{208}$Tl)~$\leq$~0.3~$\mu$Bq/kg
\end{itemize}

Given a bulk contamination in the scintillator of $\mathcal{A}$($^{208}$Tl)~$=$~0.13~$\mu$Bq/kg, the expected background level for BiPo-1, calculated by Monte Carlo simulations, is equal to 0.003 {\it back-to-back} BiPo events/month/module. For a larger BiPo detector with a sensitive surface of 12~m$^2$, this background corresponds to 0.9 BiPo-like events/month, which is low enough to reach the radiopurity performance required by the SuperNEMO experiment. 

\subsection{Measurement of the scintillators surface radiopurity}
\label{sec:surf-bkg}

Thirteen BiPo-1 modules have been used for background measurements of the scintillator surfaces. The scintillators were placed face-to-face without a foil between them. 
After 488 days of data collection, a total of 42 {\it back-to-back} BiPo events have been observed. One module appeared to be more polluted with 12 events detected in this  module. The other events were uniformly distributed in the other 12 modules and in arrival time. The contaminated module has been removed from the analysis. For the 12 remaining modules, corresponding to an integrated scintillator surface of 468~m$^2$.days, only 30 {\it back-to-back} BiPo events have been observed. 
For the same period of 488 days of data collection, the expected number of random coincidences is about 0.2  (see section~\ref{sec:random-coinc}) and the expected background due to a bulk contamination in the scintillators has a maximum of 0.6 BiPo events (see section~\ref{sec:bulk-bkg}).
Thus, the background observed in BiPo-1 corresponds to a bismuth contamination on the surface of the scintillators.
Taking into account the 27\% efficiency, calculated by simulations, to detect the BiPo cascade from a bismuth pollution on the surface of the scintillators (see Table~\ref{tab:efficiency}) with a 20\% systematic error, this corresponds to a surface background of the BiPo-1 prototype of $\mathcal{A}$($^{208}$Tl)~$= 1.5 \pm 0.3 (stat) \pm 0.3 (syst)$~$\mu$Bq/m$^2$ in $^{208}$Tl.
 
\begin{figure}[htb]
  \centering
  \includegraphics[scale=0.25, angle=270]{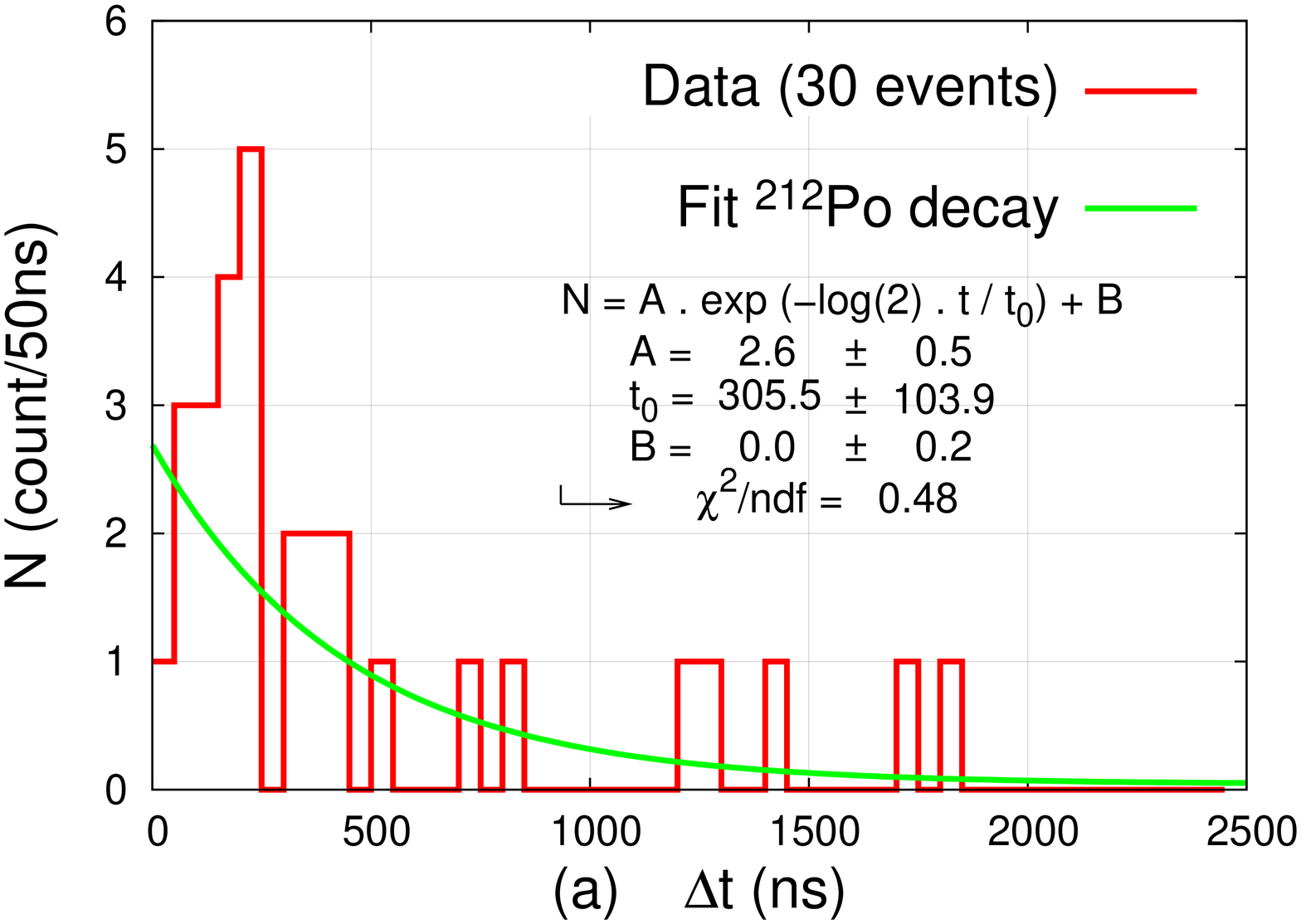}
  \hspace{0.0cm}
  \includegraphics[scale=0.25, angle=270]{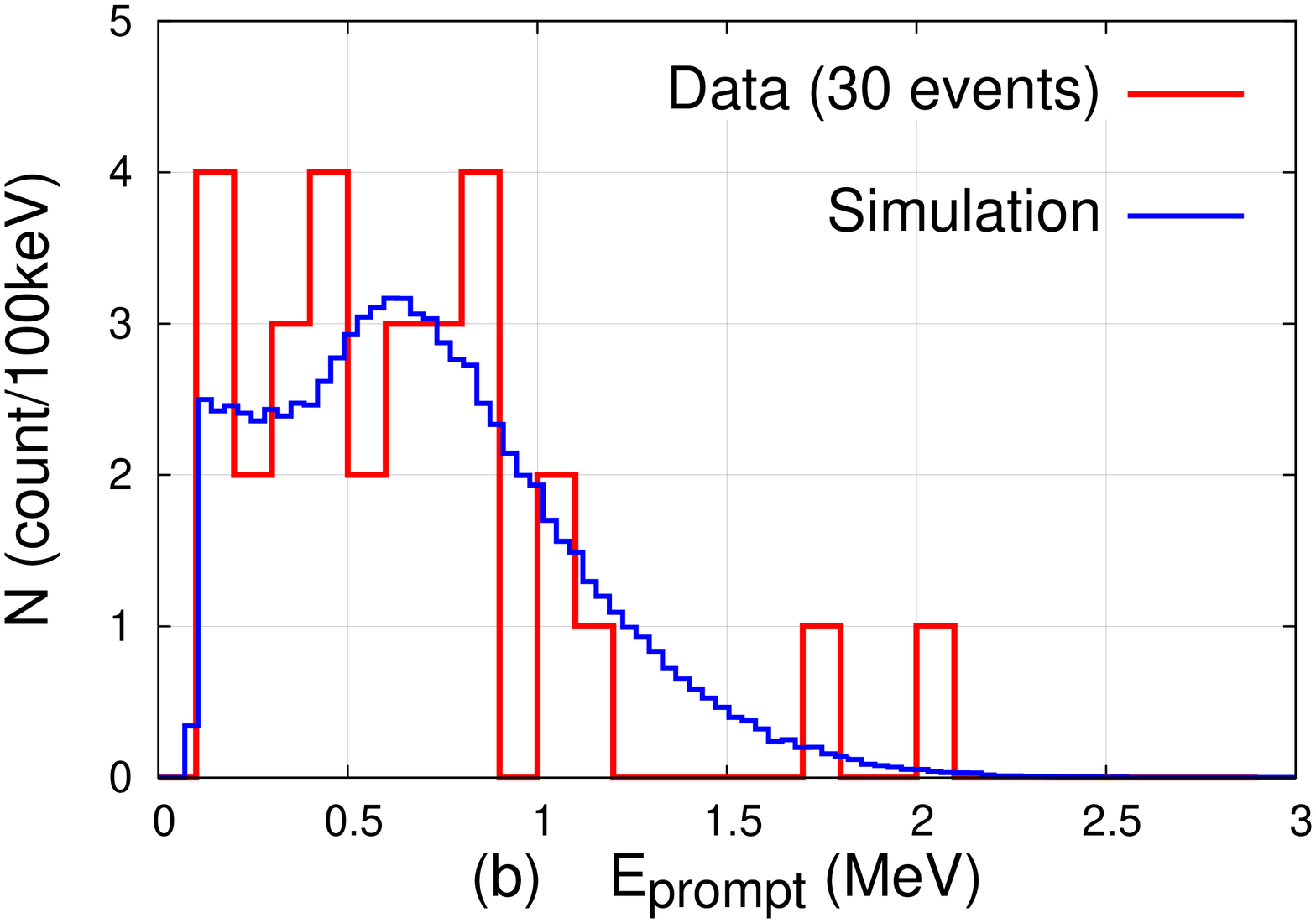}
  \hspace{0.0cm}
  \includegraphics[scale=0.25, angle=270]{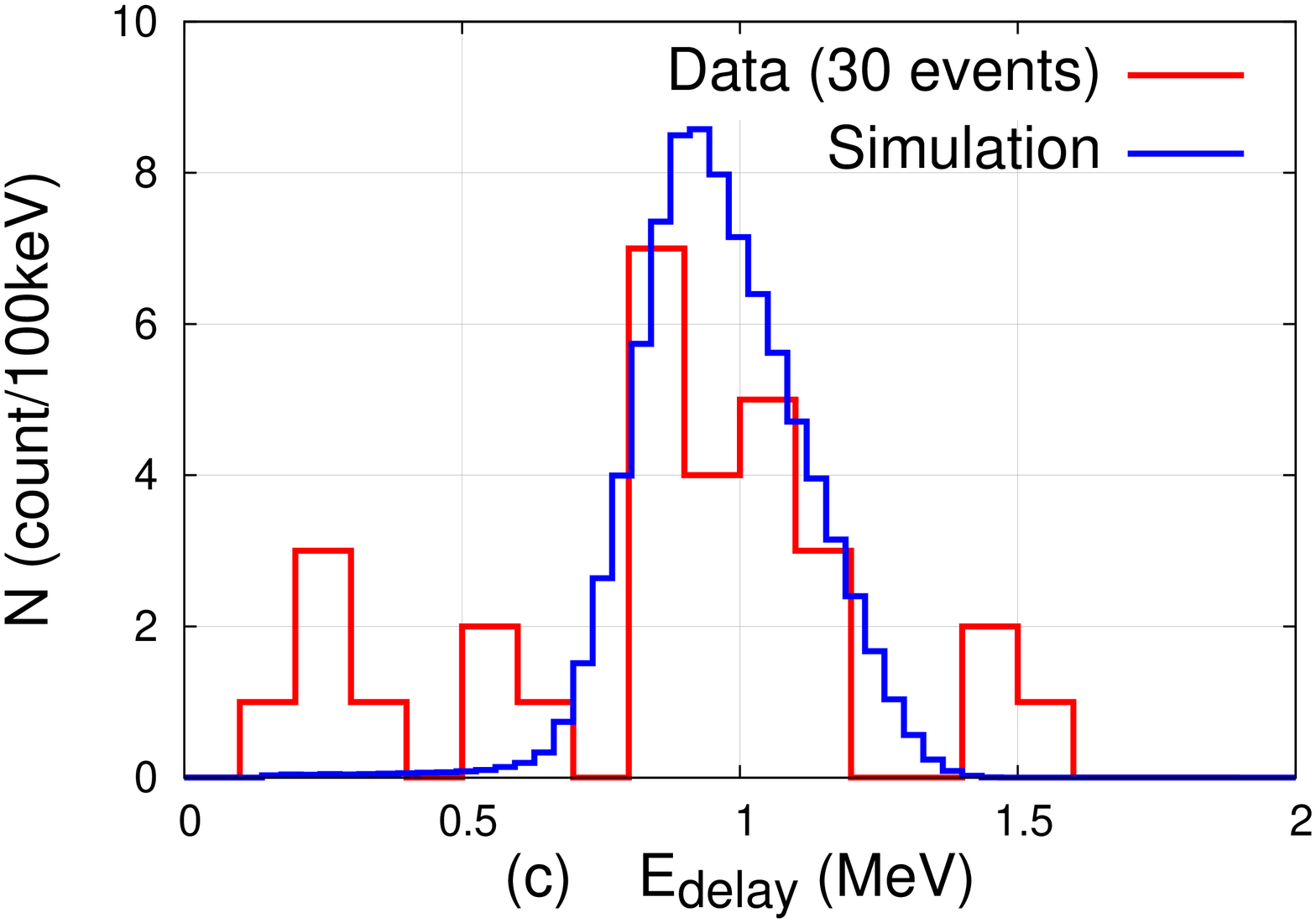}
  \caption{Distributions (a) of the delay time, (b) of the energy of the prompt signal and (c) the energy of the delayed signal, shown in red for the 30 BiPo background events observed during 423 days of data collected with 12 BiPo-1 modules, and in blue for the Monte-Carlo simulations. The green curve in the delay time distribution is the results of the exponential decay fit.}
  \label{fig:bipo-surfbkg} 
\end{figure}

The distribution of the delay time between the two signals is presented in Figure~\ref{fig:bipo-surfbkg}. Despite the low statistics, it is compatible with an exponential decay distribution with a half-life of T$_{1/2}=305 \pm 104 (stat)$~ns. 
The energy spectra of the prompt $\beta$ and delayed $\alpha$ signals, also presented in Figure~\ref{fig:bipo-surfbkg}, are in good agreement with the expected spectra calculated by the simulations. 
The energy distribution of the delayed signal is centered around 1 MeV as expected for the 8.78~MeV $\alpha$ emitted from $^{212}$Po on the surface of the scintillators with a quenching factor of 9. The origin of the few events at lower energies is still unknown.

There are several possible origins of the surface background observed with BiPo-1. 
It might be due to contaminations either during the surface machining of the scintillator plates, during the cleaning procedure of the scintillators or during the evaporation of the aluminium on the entrance face of the scintillators\footnote{The evaporation was done with a sputtering setup which was carefully cleaned.}. 
A possible contamination of the deposited aluminium is unlikely to be the origin of the background. 
The radiopurity of this aluminium has been measured with low background HPGe and $\mathcal{A}$($^{208}$Tl)~$<$~0.2~mBq/kg. This radiopurity corresponds to less than 2 BiPo events, which is much less than the 30 background events observed in BiPo-1. 
Another possible origin of the observed background is a thoron contamination between the two scintillators. Considering a typical gap of air of 200~$\mu$m, the 30 observed background events would correspond to a thoron activity of the air of $\mathcal{A}$(thoron)~$\sim$~30~mBq/m$^3$. However, this level of contamination seems too large compared to an estimation of thoron emanation from the PMTs. 
  
\begin{figure}[htb]
  \centering
  \includegraphics[scale=0.5]{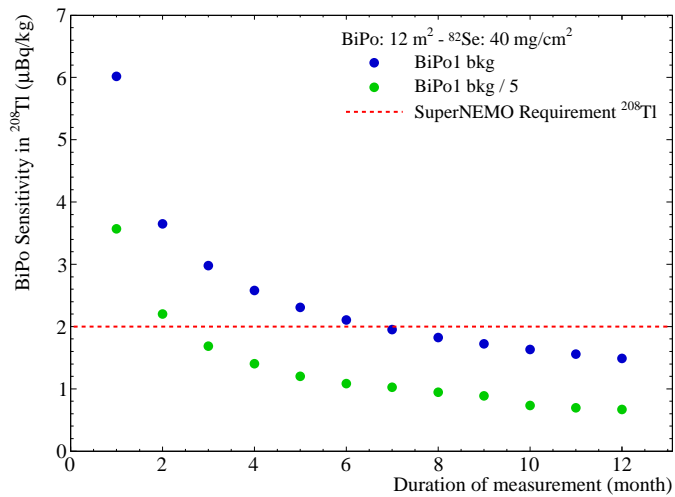}
  \caption{Expected sensitivity of a 12~m$^2$ BiPo detector in $^{208}$Tl as a function of the measurement duration (in months). Calculation is done with a 12~m$^2$ selenium foil of 40~mg/cm$^2$. Blue dots correspond to the case of a surface background of the detector of $\mathcal{A}$($^{208}$Tl)~$=$~1.5~$\mu$Bq/m$^2$, as in BiPo-1; green dots correspond to the case of a surface background five times smaller.}
  \label{fig:bipo-sensitivity-surfbkg}
\end{figure}

The result of the background measured in BiPo-1 can be extrapolated to a larger BiPo detector. 
Figure~\ref{fig:bipo-sensitivity-surfbkg} shows the expected sensitivity of the final BiPo detector in $^{208}$Tl as a function of the duration of the measurement. The calculation is done with a 12~m$^2$ selenium foil (40~mg/cm$^2$ thick) assuming that the surface background of this detector is  the same as that measured in BiPo-1 ($\mathcal{A}$($^{208}$Tl)~$=$~1.5~$\mu$Bq/m$^2$). 
The expected number of background events is equal to 17 counts/month. 
The required SuperNEMO sensitivity of 2~$\mu$Bq/kg (90\% C.L.) in $^{208}$Tl could be obtained after six months of measurement. If the level of background is five times smaller, the required sensitivity of 2~$\mu$Bq/kg can be reached after two months.

An analysis has also been performed by selecting the {\it same-side} BiPo events with prompt and delayed signals in the same scintillator in order to measure the background in this topology. A total of 223 {\it same-side} BiPo background events have been observed in Phase 1. It corresponds to a background about 7 times greater than the one measured in the {\it back-to-back} topology. The energy spectrum of the delayed signals is flat up to 1~MeV which rejects the hypothesis of contamination in the scintillator. The most probable origin of this background is a $^{212}$Bi ($^{208}$Tl) contamination in the epoxy (Araldite 2020) which has been used to attach the scintillators to the optical guide. Taking into account the uncertainty of the thickness of the epoxy, this would correspond to an activity in $^{208}$Tl of the Araldite 2020 of $\mathcal{A}$($^{208}$Tl)~$=500 \pm 300$~$\mu$Bq/kg. Recently a UV optical glue (Dymax) has been selected for future BiPo developments. Its radiopurity has been measured by HPGe spectroscopy and is $\mathcal{A}$($^{208}$Tl)~$<$~72~$\mu$Bq/kg (90\%~C.L.). The use of this glue for the next BiPo detectors might reduce strongly the background observed in the {\it same-side} topology and thus might increase the BiPo efficiency by a factor $\approx$ 1.5 by detecting both the {\it back-to-back} and {\it same-side} BiPo events.

\section*{Conclusion}
\label{sec:conclusion}

The SuperNEMO collaboration is developing a BiPo detector using thin plastic scintillator plates in order to measure with very high sensitivity the radiopurity in $^{208}$Tl and $^{214}$Bi of the double beta decay source foils to be used in the SuperNEMO experiment.

The BiPo-1 with 0.8~m$^2$ of sensitive surface has been fully operational since May 2008 in the Modane Underground Laboratory. This prototype has confirmed the proposed BiPo technique. First, the detection efficiency has been experimentaly verified by measuring a calibrated aluminium foil. Second, the different components of the BiPo-1 background have been measured. The most challenging one is a contamination on the surface of the scintillators in contact with the foils. After more than one year of data collection, a surface activity of $\mathcal{A}$($^{208}$Tl)~$= 1.5 \pm 0.3 (stat) \pm 0.3 (syst)$~$\mu$Bq/m$^2$ has been measured. 
Given this level of background, a larger BiPo detector having 12~m$^2$ of active surface area, is able to qualify the radiopurity of the SuperNEMO selenium double beta decay foils with the required sensitivity of $\mathcal{A}$($^{208}$Tl)~$<$~2~$\mu$Bq/kg (90\% C.L.) with a six month measurement.

The construction of a medium-size BiPo-3 detector with a 3.25~m$^2$ sensitive surface and using the same techniques developed in the BiPo-1 prototype has been initiated.  The goal of the BiPo-3 detector is to measure the first double beta decay source foils of the SuperNEMO demonstrator in the year 2011.

\section*{Acknowledgments}

The authors would like to thank the Modane Underground Laboratory staff for their technical support in running BiPo-1, and the IN2P3 Computing Center in Lyon for its software and computing support. This work was supported by the French Grant ANR-06-BLAN-0299 funded by the Agence Nationale de la Recherche, by the Spanish MICINN for the FPA2007-62833, FPA2008-03456 and FPA2006-12120-C03 contracts, part of which comes from FEDER funds, by the Russian RFBR 09-02-00737 Grant and by the UK STFC.



\end{document}